\documentclass[apl,showpacs,superscriptaddress,amsmath,amssymb,twocolumn]{revtex4-1}

\usepackage{graphicx}
\usepackage{epsfig}
\usepackage{rotating}
\usepackage{multirow}
\usepackage{booktabs}
\usepackage{mathrsfs}
\usepackage{tabls}
 \usepackage{longtable}
 \usepackage{tabularx}
 \usepackage{ltablex}
 \usepackage{ltxtable}
\usepackage{siunitx} %
\usepackage[flushleft, online]{threeparttablex}
\usepackage{subfigure} 

\begin{document}

\title{Band structure and giant Stark effect in two-dimensional transition-metal dichalcogenides}

\author{M. Javaid}
\email{maria.javaid@rmit.edu.au}
\affiliation{Chemical and Quantum Physics, School of Science, RMIT University, Melbourne VIC 3001, Australia}
\affiliation{The Australian Research Council Centre of Excellence for Nanoscale BioPhotonics, School of Science, RMIT University, Melbourne, VIC 3001, Australia}

\author{Salvy P. Russo}
\affiliation{Chemical and Quantum Physics, School of Science, RMIT University, Melbourne VIC 3001, Australia}
\affiliation{ARC Centre of Excellence in Exciton Science, School of Science, RMIT University, Melbourne, VIC 3001, Australia}

\author{K. Kalantar-Zadeh}
\affiliation{School of Engineering, RMIT University,
Melbourne, VIC 3001, Australia}

\author{Andrew D. Greentree}
\affiliation{Chemical and Quantum Physics, School of Science, RMIT University, Melbourne VIC 3001, Australia}
\affiliation{The Australian Research Council Centre of Excellence for Nanoscale BioPhotonics, School of Science, RMIT University, Melbourne, VIC 3001, Australia}

\author{Daniel W. Drumm}
\affiliation{The Australian Research Council Centre of Excellence for Nanoscale BioPhotonics, School of Science, RMIT University, Melbourne, VIC 3001, Australia}

\date{\today}

\begin{abstract}
We present a comprehensive study of the electronic structures of 192 configurations of 39 stable, layered, transition-metal dichalcogenides using density-functional theory. We show detailed investigations of their monolayer, bilayer, and trilayer structures' valence-band maxima, conduction-band minima, and band gap responses to transverse electric fields. We also report the critical fields where semiconductor-to-metal phase transitions occur. Our results show that band gap engineering by applying electric fields can be an effective strategy to modulate the electronic properties of transition-metal dichalcogenides for next-generation device applications.
\end{abstract}
\maketitle

\section{Introduction}

Recent advances in the development of atomically-thin materials have opened up new possibilities to explore a two-dimensional (2D) semiconducting era \cite{Novoselove, EmergingPL, Geim, 2DMat4, Golberg2010, Osada2009, 2DMat5, 2DMat1}. Among 2D materials, transition-metal dichalcogenides (TMDCs) are an interesting family with a diverse range of material properties varying from metals to insulators \cite{TMDCs_photonics}. TMDCs have general formula MX$_{2}$, where M~$\in\left\{{\rm Mo, W, Hf, Zr, Sc,}~\text{etc.} \right\}$ is a transition metal and X~$\in\left\{{\rm O, S, Se, Te} \right\}$ is a chalcogen. (In this article, we include oxygen in the chalcogens for brevity of expression.) In MX$_{2}$, there is a layer of metal atoms sandwiched between two layers of chalcogen atoms in a X--M--X pattern \cite{GeneralFormula,XLF2012}.

Many of the semiconducting, layered TMDCs have similar general features. Their band gaps widen with decreasing number of layers. For example, by reducing the number of layers from bulk down to a single layer, the band gap of MoS$_{2}$ switches from an indirect band gap of 1.28 eV to a direct band gap of 1.80 eV \cite{PRL105,EmergingPL} due to the stronger quantum confinement in the vertical direction. This band gap tunability via the layer thickness provides further avenues for novel nanoelectronics and nanophotonics applications \cite{PRL105,EmergingPL}. 

A tunable band gap is highly desirable to allow design flexibility and to control the properties of electronic devices, \textit{e.g.}, single-electron transistors \cite{SET_Angus, SET, Hongqi_2}, to achieve tunable fluorescence, display technology, and in changing electrical conductivity applications \textit{etc.} \cite{Wiley}. However, after fabrication the layer thickness cannot be varied to further tune the band gap. One possible way to control the device properties after the fabrication process is to tune the band structure using external transverse electric fields \cite{Ning2014, Dongwei, Balu2012}. 

Two-dimensional semiconductors are also good candidates for photocatalytic water applications due to their large specific surface area, excellent light absorption, and tunable electronic properties \cite{Yeh2014, Singh2015}. It has been reported in \cite{RT2015} that many of the TMDCs qualify for water-splitting applications. 

In the present work, our goal is to explore the band structure modification via electric field of as many stable, layered, predominately semiconducting TMDCs as possible. Around 40 of the layered TMDCs were reported by Wilson \textit{et al.} in the 1960s \cite{Wilson_1960} and reviewed recently by Kuc \textit{et al.} \cite{LayeredTMDC}. They reported the bulk structures and electrical characteristics of the H and/or T phases of ScS$_{2}$, ScSe$_{2}$, ScTe$_{2}$, TiS$_{2}$, TiSe$_{2}$, TiTe$_{2}$, ZrS$_{2}$, HfS$_{2}$, HfSe$_{2}$, VS$_{2}$, VSe$_{2}$, VTe$_{2}$, NbS$_{2}$, NbSe$_{2}$, NbTe$_{2}$, TaS$_{2}$, TaSe$_{2}$, TaTe$_{2}$, CrS$_{2}$, CrSe$_{2}$, CrTe$_{2}$, MoS$_{2}$, MoSe$_{2}$, MoTe$_{2}$, WS$_{2}$, WSe$_{2}$, WTe$_{2}$, MnS$_{2}$, MnSe$_{2}$, MnTe$_{2}$, ReS$_{2}$, ReSe$_{2}$, ReTe$_{2}$, FeS$_{2}$, FeSe$_{2}$, FeTe$_{2}$, NiS$_{2}$, NiSe$_{2}$, NiTe$_{2}$, and PdS$_{2}$. There have been significant efforts to synthesize single and multilayer nanosheets from bulk structures, \textit{e.g.}, TiS$_{2}$, ZrS$_{2}$, NbSe$_{2}$, TaS$_{2}$, TaSe$_{2}$, MoS$_{2}$, MoSe$_{2}$, MoTe$_{2}$, WS$_{2}$, WSe$_{2}$, WTe$_{2}$, ReS$_{2}$, ReSe$_{2}$, and NiTe$_{2}$ have been grown by various methods \cite{Miguel, Miguel2014, Barja2016, Dumceno, Xingli, Chang2014, Chen2015, Yoshida2016, CHUNG1998137, Keyshar2015, Hafeez2016, Naylor2016, Zhou2017, Dumitru, Eichfeld, Gao2015, Coleman568, Zeng2011, Zhan, C2NR31833D, ENNAOUI1995124, Kang2013, Zhang2013, Shaw2014, Hsien2012} and discussed in recent reviews of TMDCs \cite{ChemTMDC,TMDCReview1,TMDCsReview}. Our current work also explores those TMDC monolayers and multilayers which have not yet been synthesized but have been predicted to be stable, layered, and semiconducting.

Molybdenum- and tungsten-based dichalcogenides have been intensively investigated. The band gap variation of a few layers of these materials under electric field have been widely studied \cite{ RNT2011, LLGC2012, Nguyen2016, LayeredTMDC, MoS2_ML_EF, WS2BL, WSe2BL_field, WSe2MultiL}. Leb\`{e}gue \textit{et al.} \cite{bulk_TMDC} have reported 2D materials by data filtering from their known bulk structures. The band gaps of monolayers of the stable, layered TMDCs have been reported by Ataca \textit{et al.} in \cite{stabilityPaper} and later by Rasmussen \textit{et al.} in \cite{RT2015}. However, to date there has been no comprehensive study of the band structure responses of the stable, semiconducting, few-layer TMDCs to electric fields. 

In this article, we study the role of transverse electric field in engineering band gaps in the known, stable, predominately semiconducting, few-layer TMDCs. We also study responses of their valence-band maxima (VBM) and the conduction-band minima (CBM) with electric field. This is of especial significance for device design and the optimization of atomically thin optoelectronic systems \cite{Wiley, Hongqi_1, SET_Angus}.

\begin{figure}[bt]
\centering
\includegraphics[width=0.99\linewidth]{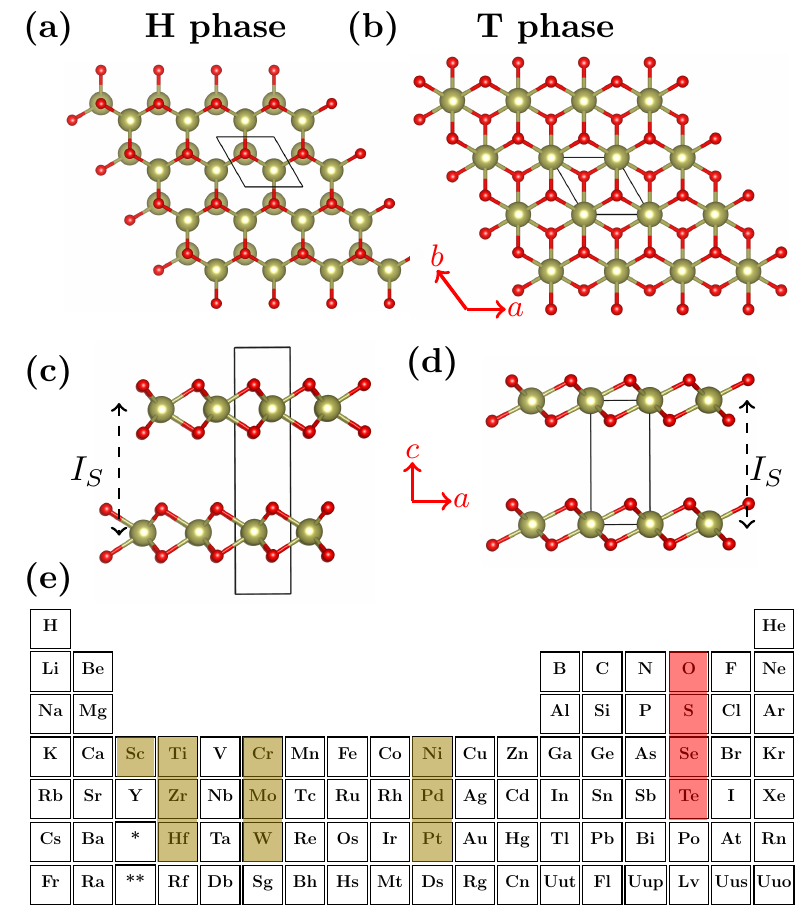}
\caption{Bulk TMDC structures; (a) top view of the H phase, (b) top view of the T phase, (c) side view of the H phase, (d) side view of the T phase. Large, brown circles are metal atoms and small, red circles are chalcogens. The unit cells have lengths $a = b$ and $c$ along the directions shown by axes labels between the corresponding sub-figures. The interlayer separation $I_{S}$ (from centre to centre of the metal atoms) between the two TMDC layers is shown by dashed, arrowed lines. (e) Periodic table (without lanthanides and actinides) showing the brown-highlighted transition metals and red-highlighted chalcogens which we have studied in this article.}
\label{Fig:Phases}
\end{figure}

This paper is organized as follows: In Section II, we describe the computational details. Then we discuss the electronic structures in Sec.~III, followed by the lattice parameters and the band gap analysis at zero field in Sec.~III A. Then we discuss the variations of the band gaps under field for monolayer, bilayer, and trilayer TMDCs in Sec.~III B followed by the discussion of the responses of the VBM and CBM to the electric field in Sec.~III C.
\section{Computational details}

We studied TMDCs constructed by combining transition metals (brown-highlighted) and chalcogens (red-highlighted) in the periodic table as shown in Fig.~\ref{Fig:Phases}(e). We studied two different structures: honeycomb (H) crystal structures with hexagonal space group P6$_{3}$/mmc and centred honeycomb (T) crystal structures with trigonal space group P$\overline{3}$m1. Both of these structures and their unit cells are shown in Fig.~\ref{Fig:Phases}.

All the calculations were performed using zero-Kelvin, density-functional theory (DFT) in the \textsc{crystal09} code \cite{dovesi2005crystal, Dovesi2013}. The lattice constants of the bulk structures were determined by structure relaxation using the PBEsol functional \cite{Perdew2008} for both exchange and correlation terms. The PBEsol functional generally predicts lattice constants more accurately than PBE and LSDA (thus better approximating the equilibrium properties of solids), and also handles the electronic response to potentials better than most GGA functionals \cite{Perdew2008}. We compared our computed $c$ lattice parameter with the experimental values where available. We did not find a difference larger than 5\% between the experimental and our computed values, indicating that the PBEsol functional is reasonably calculating this parameter. As there are no published van der Waals correction factors for third-row transition metals \cite{Grimme}, we ignored these corrections to keep our calculations consistent for all studied materials. 

The geometries were optimized to the default \textsc{crystal09} convergence criteria of less than 4.5$\times 10^{-4}$ Hartree/Bohr maximum force, 3$\times 10^{-4}$ Hartree/Bohr RMS force, 1.8$\times 10^{-3}$ Hartree maximum displacement, and 1.2$\times 10^{-3}$ Hartree RMS displacement. 

We used Gaussian basis sets; triple-zeta for valence electrons plus a polarization function (TZVP) for the lighter elements (third-period metals and all chalcogens but Te) and pseudopotential basis sets for the heavy elements as follows: Sc\_pob\_TZVP\_2012 \cite{Sc_BS} for Sc, Ti\_pob\_TZVP\_2012 for Ti \cite{Sc_BS}, Zr\_ECP\_HAYWSC\_311d31G\_dovesi\_1998 \cite{Zr_BS} for Zr, Hf\_ECP\_Stevens\_411d31G\_munoz\_2007 \cite{Hf_BS} for Hf, Cr\_pob\_TZVP\_2012 \cite{Sc_BS} for Cr, Mo\_SC\_HAYWSC-311(d31)G\_cora\_1997 \cite{Mo_BS} for Mo, W\_cora\_1996 \cite{W_BS} for W, Ni\_pob\_TZVP\_2012 \cite{Sc_BS} for Ni, Pd\_HAYWSC-2111d31\_kokalj\_1998\_unpub \cite{Pd_BS} for Pd, Pt\_doll\_2004 \cite{Pt_BS} for Pt, O\_pob\_TZVP\_2012 \cite{Sc_BS} for O, S\_pob\_TZVP\_2012 \cite{Sc_BS} for S, Se\_pob\_TZVP\_2012 \cite{Sc_BS} for Se, and Te\_m-pVDZ-PP\_Heyd\_2005 \cite{Te_BS} for Te.  

\begin{figure*}[tb!]
\hspace*{-0.2 in}
\centering
\includegraphics[scale=1.3]{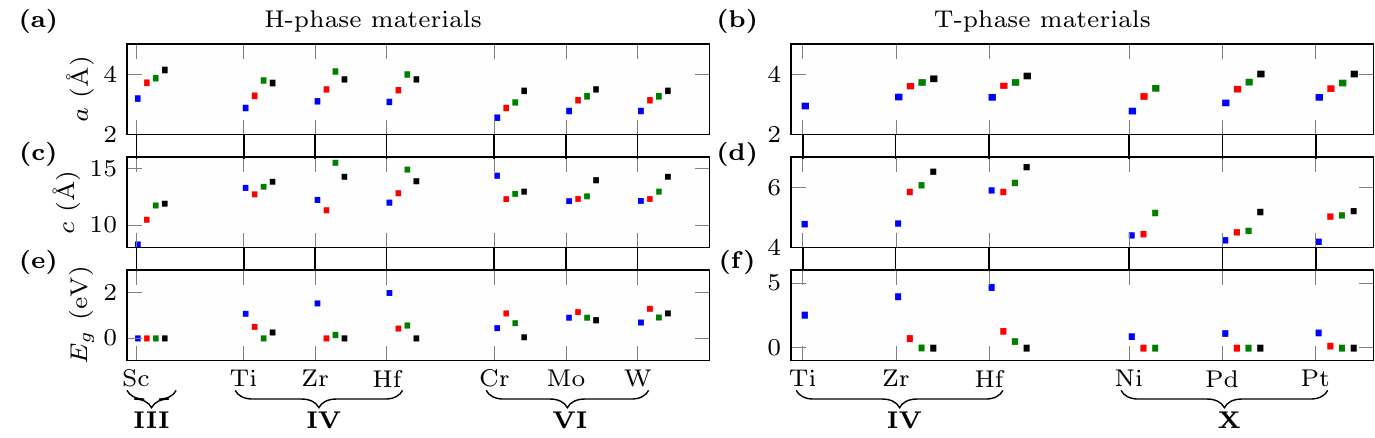}
\caption{Bulk structures' in-plane lattice parameters; $a$ for (a) H-phase and (b) T-phase, perpendicular to the plane, $c$ for (c) H-phase and (d) T-phase. Bulk structure band gaps for (e) H-phase and (f) T-phase materials. The colors of the rectangles represent the chalcogens; blue for oxides, red for sulphides, green for selenides, and black for tellurides. Transition metal group numbers are indicated below the braces.}
\label{Fig:LParameters}
\end{figure*}

We created the monolayer, bilayer, and trilayer TMDC unit slabs by defining (001) planes from bulk models, and including vacuum to a total cell height of $c = 500$~\AA. We set an 8$\times$16$\times$1 Monkhorst-Pack \cite{Monkhorst1976} $k$-point mesh. We optimized the geometries of monolayer, bilayer, and trilayer TMDC unit slabs under zero electric field. We used these zero-field-optimized geometries to study the effects of transverse electric fields on their band structures as the influence of field-based geometric disturbances on the band structures is negligible \cite{LLGC2012}. We also checked the effects of electric field on the interlayer separation in H-TiO$_{2}$ and H-MoS$_{2}$ bilayer structures and found that the electric fields used do not modify the optimal interlayer separations of these structures. 

The applied field strength was consistently varied from 0 to $\pm$ 0.2 V/\r{A} as discussed later in the relevant sections. To compute the critical field (where the semiconductor-to-metal phase transition occurs), we further increased the field strength above 0.2 V/\r{A} until we reached the field value where the band gap closed, except for those materials that have critical fields smaller than 0.2 V/\r{A}. 

We calculated the band structures along the high-symmetry path \textbf{$ \boldsymbol{\Gamma}$-M-K-$\boldsymbol{\Gamma}$}. Band structures were calculated for uniformly varying electric fields applied perpendicular to the TMDC slabs.

\section{Results and discussion}

\begin{figure*}[hbt]
\centering
\includegraphics[width=0.99\linewidth]{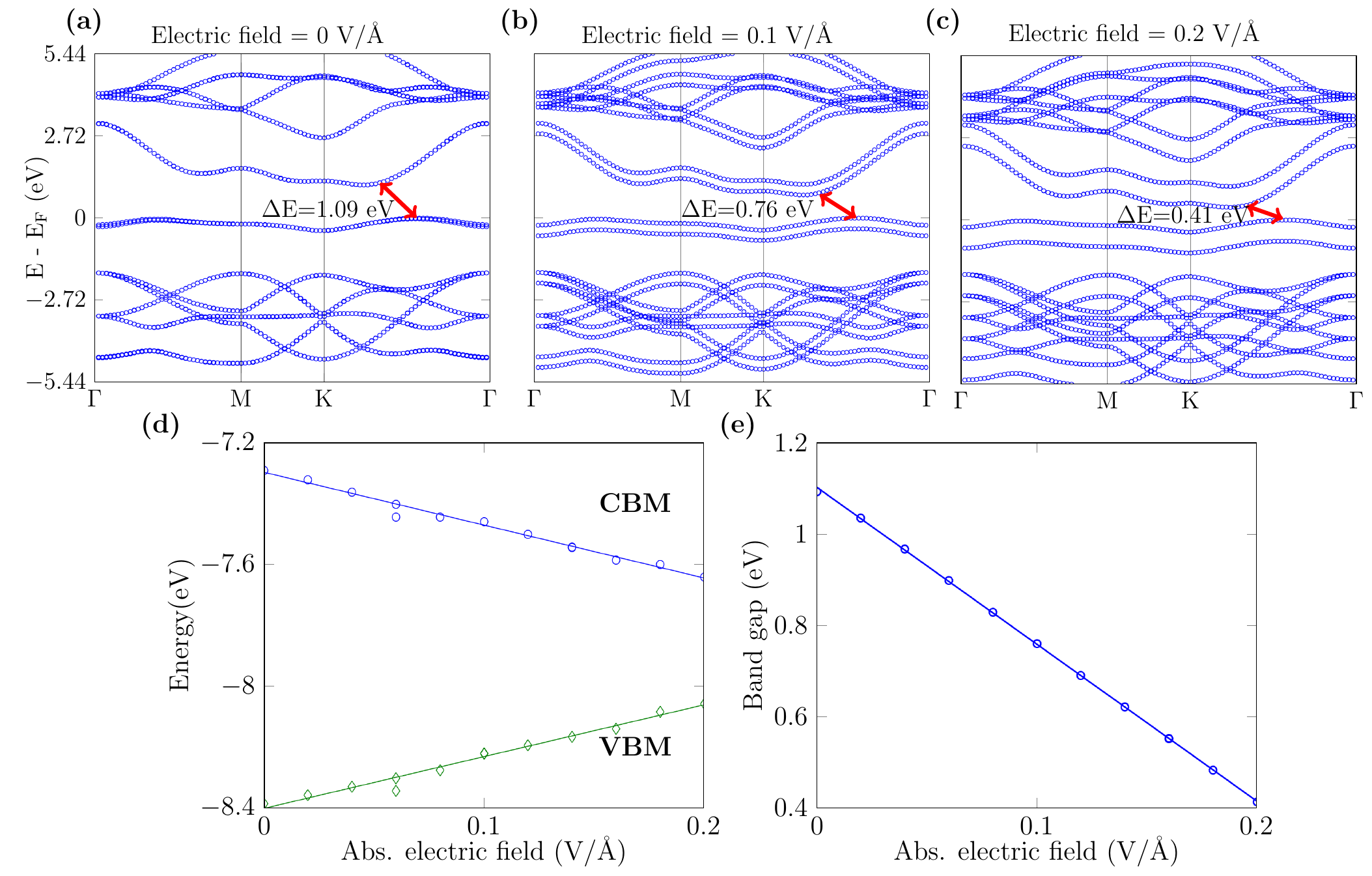}
\caption{H-TiO$_{2}$ bilayer sample band structures with zero and finite field. Band structures of: (a) H-TiO$_{2}$ bilayer at zero field with an indirect band gap of 1.09 eV (red-arrowed lines) from the CBM to the VBM, (b) H-TiO$_{2}$ bilayer at an electric field of 0.1 V/\r{A} with an indirect band gap of 0.76 eV, (c) H-TiO$_{2}$ bilayer at an electric field of 0.2 V/\r{A} with an indirect band gap of 0.41 eV. (d) Modulation of the VBM and CBM of the H-TiO$_{2}$ bilayer for various values of absolute electric field (Note, the energy zero here is the energy of the vacuum.). (e) Band gap of the H-TiO$_{2}$ bilayer for various values of the absolute electric field showing a decrease in the band gap with the field.}
\label{Fig:Sample}
\end{figure*}

\begin{figure*}[p!]
\centering
\includegraphics[width=0.99\linewidth]{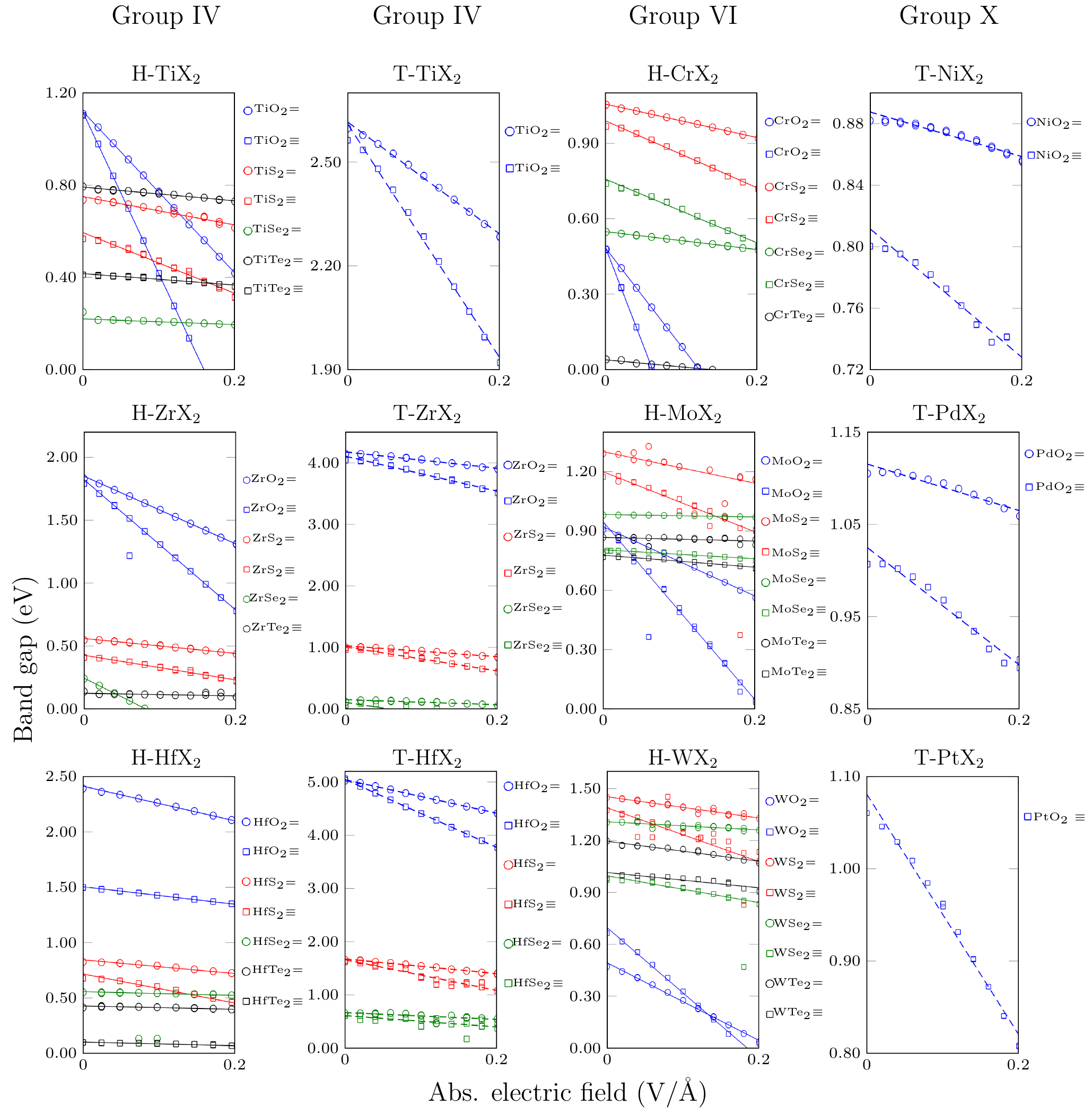}
\caption{Band gaps as functions of absolute electric field strength for all studied H- and T-phase, bilayer and trilayer TMDCs (except the Sc family, which are all metallic - see Table 1). The subfigures separate the gamut into families of common phase and metal element, and are arranged preserving period and group order. In each subfigure, the colors denote the chalcogens; blue for oxides, red for sulphides, green for selenides, and black for tellurides. Bilayers are represented by circular markers (and = symbols in the legend for brevity), and trilayers by square markers (and $\equiv$ symbols in the legend). The lines (solid for H phase and dashed for T phase) are linear fits to each data set, excluding any inconsistent points (as discussed in the text). Most materials exhibit monotonic band gap decrease with increasing electric field, at least to their critical fields. Note that zero-band-gap and zero-band-gap-modulation values are not shown for clarity, and that the vertical scales differ between subfigures. }
\label{Fig:BG_Field}
\end{figure*}

In this section, we present our electronic structure calculations. For all the TMDC materials under investigation, we computed the relaxed bulk (three-dimensional) structure parameters (the in-plane lattice parameters or unit-cell lengths ($a$) and the unit-cell lengths ($c$) perpendicular to the plane), the interlayer separations $I_{S}$, and band gaps of the bulk structures without electric field. We then computed the band gaps of the relaxed monolayer, bilayer, and trilayer structures with and without electric field along the $\pm c$ directions. We discuss the band gaps' modulation, the critical fields where the semiconductor-to-metal phase transition occurs and the responses of the VBMs and CBMs under electric field. We report our computed parameters in Table \ref{Tab: Table1} (which is at the end of the document due to its considerable length). We also report the published values of the monolayer, bilayer, and trilayer structures' band gaps at zero field, band gap variations under field, and  the critical field where available. Throughout the discussion, we focus more on the trends observed across the TMDC family, in various ways, instead of the in-depth analysis of individual materials. 

\subsection{Properties under zero field}

For several of the materials (H-ScX$_{2}$, T-ZrX$_{2}$, T-NiX$_{2}$, T-PdX$_{2}$, and T-PtX$_{2}$), we observe increasing bulk structure lattice parameters $a$ and $c$ as we move down the chalcogen group from oxides to tellurides, accompanied by decreasing band gaps (Fig.~\ref{Fig:LParameters}), as reported in Table \ref{Tab: Table1}. The other materials depart from this trend in various ways, to various extents. For example, H-MoX$_{2}$ and H-WX$_{2}$ follow this trend for both their lattice parameters but their bulk band gaps from oxides to tellurides do not. H-CrX$_{2}$ follow an increasing trend in the lattice parameter $a$ but not in $c$. Also their bulk band gaps appear non-linear with respect to chalcogen element. T-HfX$_{2}$ follow the trend in $a$ and the bulk band gaps but T-HfS$_{2}$ deviates for $c$. H-ZrX$_{2}$, H-HfX$_{2}$ and H-TiX$_{2}$ show deviation from this general trend across all of the parameters, \textit{i.e.}, $a$, $c$, and the bulk band gaps. A similar oxide to telluride trend in TMDC monolayer band gaps has been reported by Rasmussen \textit{et al.}~\cite{RT2015}. No major trends in the lattice parameters or bulk band gaps are observed either across the periods or down the groups in the transition metals (Fig.~\ref{Fig:LParameters}). 

Companion analyses have been carried out by instead changing the transition metal period (3, 4, and 5), transition metal group (III, IV, VI, and X), the material phase (H and T), or the number of layers in the model (1, 2, or 3), while holding all other dimensions constant.  The number of non-singleton, non-zero-valued, subset classes for each dimension are: $12(\times 4)$ varying chalcogen, $12(\times 3)+4(\times 2)$ varying transition-metal period, $4(\times 3)+17(\times 2)$ varying transition-metal group, $9(\times 2)$ varying material phase, and $46(\times 3)$ varying the number of layers.  Note: the numbers in parentheses indicate the number of members of each class.

The results are summarised in Table \ref{tab:bulksum}, which shows the dominant behavior (or the two, most-prevalent, non-dominant behaviors) across all subset classes with more than one element.  The behaviors considered are: flat (F), monotonically increasing ($\uparrow$), monotonically decreasing ($\downarrow$), or other non-monotonic behavior (X) for classes with more than two members. 

\begin{table}[tb!]
\begin{ruledtabular}
\begin{tabular}{c|cccc}
\hline
Variable & \multicolumn{4}{c}{Parameter}\\
 & $a$	& $c$	& Bulk $E_{g}$	& $E_{g}(\mathscr{E}=0)$\\
\hline
Chalcogen	& $\uparrow$	& $\uparrow$/X$^{a}$	& X/$\downarrow^{b}$	& $\downarrow$\\
TM period	& $\uparrow$/X$^{c}$	& X/$\uparrow^{d}$	& $\uparrow$	& $\uparrow$\\
TM group	& $\downarrow$	& $\downarrow$	& $\uparrow$/$\downarrow^{e}$	& $\uparrow/\downarrow^{f}$\\
Phase	& $\uparrow$	& $\downarrow$	& $\uparrow$	& $\uparrow$\\
Layers	& F	& F	& F	& $\downarrow$\\
\hline
\end{tabular}
\begin{tablenotes}
{\footnotesize	
\item$^{a}$ 50/50\%; $^{b}$ 50/42\%; $^{c}$ 50/38\%; $^{d}$ 44/38\%; $^{e}$ 43/38\%; $^{f}$ 47/42\%}
\end{tablenotes}
\caption{Multidimensional analysis of dominant behaviour(s) within the TMDC family for the lattice parameters $a$ and $c$, the bulk band gap $E_{g}$, and the layer-dependent band gap $E_{g}(\mathscr{E}=0)$.  F, $\uparrow$, $\downarrow$, and X respectively indicate flat, monotonically increasing, monotonically decreasing, and other non-monotonic behaviour with an increase to the relevant dimension. The dimensions (with the ordering used in the analysis) are: chalcogen (period 2, 3, 4, or 5), transition-metal period (3, 4, or 5), transition-metal group (III, IV, VI, or X), material phase (from H to T), and number of layers (1, 2, or 3).  Single indicators are shown when $>$50\% of the subset classes exhibit the behavior; pairs of indicators are shown when there are behaviors that capture most subsets (their relative percentages are shown as footnotes).}
\label{tab:bulksum}
\end{ruledtabular}
\end{table}

Examining combinations of the trends (multi-parameter patterns), we found that with a change of phase (from H to T) 5 of 9 subsets showed a combined increase to $a$, decrease to $c$, and increasing band gap.  With increasing transition-metal group, 10 of 21 subsets showed decreases to both $a$ and $c$ (but a 4/6 split between increasing and decreasing band gap).  All other combinations accounted for smaller proportions of the subset classes and are therefore regarded as insignificant.

For most of the materials we find an increase in the band gap with the decreasing number of layers from bulk to monolayer which is consistent with the properties of the Mo- and W-based dichalcogenides \cite{MoWS2}. (Note there are too many values to show clearly in a figure, but their values are available in Table \ref{Tab: Table1}.) We observe minor deviations from this trend by H-ZrO$_{2}$, H-ZrSe$_{2}$, H-HfO$_{2}$, T-HfO$_{2}$, H-HfSe$_{2}$, H-CrS$_{2}$, H-CrSe$_{2}$, H-CrTe$_{2}$, H-MoO$_{2}$, H-MoSe$_{2}$, H-MoTe$_{2}$, H-WO$_{2}$, H-WTe$_{2}$, T-NiO$_{2}$, T-PdO$_{2}$, and T-PtO$_{2}$. For example, the band gap of H-ZrO$_{2}$ increases from 1.54 eV to 2.08 eV from bulk to its bilayer structure but the monolayer has a smaller gap of 1.62 eV. 

\subsection{Band gap variations under electric field}
\label{sec:bands}

An electric field can potentially be used to tailor the band gaps of layered materials. The band structure variations with electric field arise due to the well-known Stark effect. The Stark effect induces a potential difference between the layers which causes splitting of energy bands belonging to different layers as well as shifting of the VBM and CBM \cite{LLGC2012}. This splitting and shifting of the energy bands can increase \cite{Graphene_Field, Graphene_field2, BG_Field} or reduce \cite{CN_Field, RNT2011} the band gaps. In the case of multilayered TMDCs, this splitting and shifting pushes both the CBM  and VBM towards the Fermi level. This results in the reduction of the band gap with applied electric field in multilayered TMDCs. 

\begin{figure}[bt]
\centering
\includegraphics[width=0.99\linewidth]{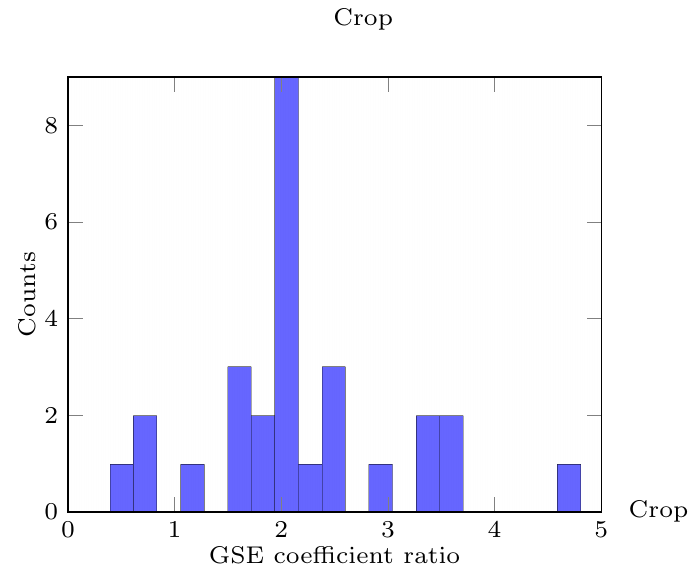}
\caption{Histogram of the trilayer to bilayer GSE coefficient ratios showing most of the materials clustered around a value of 2 which might be due to the two interlayer spacings of the trilayer as compared to the bilayer structures.} 
\label{Fig:Histogram}
\end{figure}

For each material under investigation, we computed band gaps of its monolayer, bilayer, and trilayer structures with an external electric field applied perpendicular to the layers along the $\pm c$ directions. We consistently varied the field strength from 0 to $\pm$ 0.2 V/\r{A}, in steps no larger than 0.02 V/\r{A}, except for materials (H-TiO$_{2}$ trilayer, H-ZrSe$_{2}$ bilayer, T-ZrSe$_{2}$ trilayer, H-CrO$_{2}$ bilayer and trilayer, H-CrTe$_{2}$ bilayer, and H-WO$_{2}$ trilayer) whose band gaps close before 0.2 V/\r{A}. This maximum field strength is stronger than usual device fields and is shown to more clearly illustrate the band structure variations under electric field. 

For all structures, we find that monolayer TMDCs do not respond to electric field up to a field strength of 0.2 V/\r{A}. However the band structures of most of the bilayer and trilayer TMDCs do vary with electric field offering applications in band engineering. Such results are consistent with previously reported studies of Mo- and W-based dichalcongenides. 

Figure \ref{Fig:Sample} shows sample band structures of H-phase TiO$_{2}$ bilayers at zero field and for two different finite values of the electric field. Increasing the electric field causes splittings of the energy bands resulting in a reduction of the band gap from 1.09 eV (at zero field) to 0.41 eV (at 0.2 V/\r{A}). The responses of the VBM and CBM are shown in Fig.~\ref{Fig:Sample}(d) for electric fields varying from 0 V/\r{A} to $\pm $0.2 V/\r{A}. There is significant variation of both the VBM and the CBM with the electric field pushing them towards the Fermi level, E$_{\rm F}$ for the H-TiO$_{2}$ bilayer resulting in the reduction of the band gap under field. The band gap as a function of the absolute electric field is shown in Fig.~\ref{Fig:Sample}(e) where the solid line is a linear fit to the data. For an electric field of 0.2 V/\r{A}, the band gap is reduced by 686 meV. Thus band gap variation in H-TiO$_{2}$ bilayers is achievable via electric field at a rate of -3.43 eV/(V/\r{A}) while the corresponding VBM and CBM bendings are 1.69 and -1.73 eV/(V/\r{A}) respectively.

\begin{figure*}[hbt]
\centering
\includegraphics[width=0.99\linewidth]{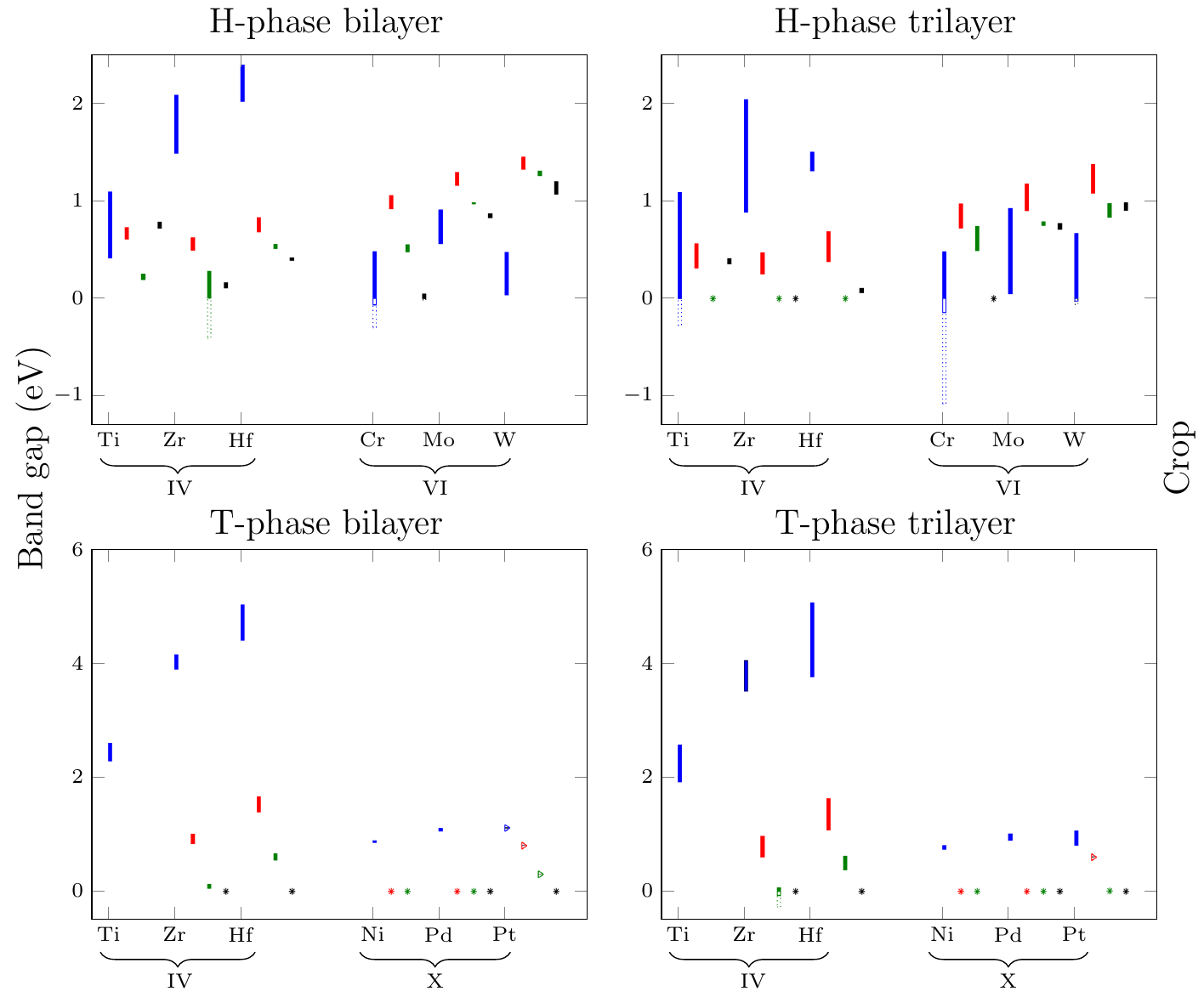}
\caption{Band gap variations with electric field strength in H- and T-phase bilayer and trilayer TMDCs. The top of each rectangular bar shows the band gap at zero field while its bottom shows the band gap at a maximum field of 0.2 V/\r{A} or less for materials whose band gaps close before 0.2 V/\r{A} as discussed in the main body and in Table III. These are shown as unfilled rectangles crossing the zero-band-gap line as far as has been calculated. To allow comparisons with non-metallic TMDCs, dotted rectangles extrapolate the pre-critical-field data to 0.2 V/\r{A} field to allow rapid, direct comparison with the other (solid) rectangles as regards the pre-critical-field behavior. This is directly related to the giant Stark effect coefficient, $\hat S$, as explained in the main body. The color of the bars represents the associated chalcogen; blue for oxides, red for sulphides, green for selenides, and black for tellurides. Metallic or semimetallic TMDCs, which have zero band gap, are denoted by asterisks. Materials whose non-zero band gaps do not change with electric field are denoted by triangles.}
\label{Fig:BandGaps}
\end{figure*}

We obtain qualitatively similar band structure responses to electric field for most of the studied materials. The band gaps as functions of electric field are shown in Fig.~\ref{Fig:BG_Field} for all materials. Most band gaps decrease monotonically with electric field strength. The Ni, Pd, and Pt oxides seem to exhibit slightly quadratic behavior at low field which is similar to the band structure modulation via electric field reported for few layer black phosphorus \cite{Phosphorus}. The linear fits applied across all band gap data as functions of electric field (Fig.~\ref{Fig:BG_Field}) are shown by solid lines for all H-phase materials and by dashed lines for all T-phase materials. Circular and square markers represent the bilayer and trilayer data in all the sub-figures while the blue, red, green, and black colors represent the oxides, sulphides, selenides, and tellurides respectively. All the materials show symmetric response to fields oriented in the $\pm c$ directions which is due to the symmetry of the chalcogen layers around the metal layer, except a few odd data points (Fig.~\ref{Fig:BG_Field}). We have excluded these odd data points from the linear fits as we believe that they do not represent the underlying physics.

The change in the band gap with the electric field, $\frac{dE_{g}}{d\mathscr{E}_{F}}$, is given by \cite{RNT2011}  
\begin{equation}
\frac{dE_{g}}{d\mathscr{E}_{F}} = -e \hat S.
\end{equation} 
Here $E_{g}$ is the band gap in eV, $\mathscr{E}_{F}$ is the electric field strength in V/\r{A}, $e$ is the fundamental charge, and $\hat S$ is the GSE coefficient, calculated from the slope of the linear fits to our data shown in Fig.~\ref{Fig:BG_Field}. We report this GSE coefficient ($\hat S$) in Table \ref{Tab: Table1} along with 95\% confidence intervals. We compare our computed GSE coefficient with the literature where available in Table \ref{Tab: Table1}.

We observe an interesting trend in $\hat S$ for the bilayer and trilayer structures of several materials. We find that the $\hat S$ values for trilayer structures are approximately twice those for bilayer structures, with a few anomalies. For example, H-TiO$_{2}$ trilayers has $\hat S$ = 6.87 \r{A} which is twice that of 3.43 \r{A} for H-TiO$_{2}$ bilayers. Similarly, H-CrO$_{2}$ and H-CrS$_{2}$ show a factor of two in $\hat S$ between their trilayer and bilayer structures. We report this trilayer-bilayer $\hat S$ ratio ($\hat S_{\equiv} / \hat S_{=}$) in Table \ref{Tab: Table1} and display it in histogram form in Fig.~\ref{Fig:Histogram}. The symbols $\equiv$ and $=$ represent trilayer and bilayer structures respectively. A group of the materials cluster around a $\hat S_{\equiv} / \hat S_{=}$ ratio of two with a few outliers. T-ZrSe$_{2}$, H-HfO$_{2}$, H-CrSe$_{2}$, H-MoO$_{2}$, H-MoSe$_{2}$, H-WSe$_{2}$, and T-NiO$_{2}$ are outliers. The tellurides show anomalous behavior where they have non-zero band gaps in both bilayer and trilayer structures. Their $\hat S_{\equiv} / \hat S_{=}$ ratios are smaller than the other dichalcogenides suggesting that telluride trilayers are less sensitive to field. 

The near-double modulation of trilayer TMDCs compared to their associated bilayers might be explained by the number of interlayer separations ($I_{S}$) they have. Bilayers have one $I_{S}$, while trilayers have two. Monolayers have none, and it is well established that they do not exhibit band gap variation (and therefore their $\hat S=0$). 

All studied materials' band gap variations with field are shown in Fig.~\ref{Fig:BandGaps}. The length of each bar shows the change in the band gap under field. The top of each rectangular bar shows the material's band gap at zero field and the bottom of each rectangular bar shows the band gap at the maximum field of 0.2 V/\r{A} or less for materials whose band gaps close before 0.2 V/\r{A}. Blue, red, green, and black colors represent the oxides, sulphides, selenides, and tellurides. Rectangular bars crossing the zero-band-gap line indicate materials where the CBM switches below VBM at maximum field.  

Table \ref{tab:summary} details a multidimensional trend analysis for $\hat{S}$ (and several other properties which are discussed below), similar to those for bulk properties presented in Table \ref{tab:bulksum}.  The numbers of subset classes change due to the separation of materials by number of layers and subsequent exclusion of monolayer data or others whose $\hat{S}$ are uniformly zero, along with any completely metallic or semi-metallic classes.  For $\hat{S}$ (and $\mathscr{E}_{F_{c}}$, $\partial E_{{\rm VBM}}/\partial E$, and $\partial E_{{\rm CBM}}/\partial E$, discussed below), there are $19(\times4)+2(\times3)$ subset classes varying chalcogen, $19(\times 3)+2(\times2)$ classes varying transition-metal period, $8(\times 3)+30(\times 2)$ varying transition-metal group, $17(\times 2)$ varying phase, and $34(\times 2)$ varying layer number.

\begin{table*}[tb!]
\begin{ruledtabular}
\begin{tabular}{c|cccccccc}
\hline
Variable & \multicolumn{8}{c}{Parameter}\\
& $\hat{S}$	&	$\mathscr{E}_{F_{c}}$ &  $E_{{\rm VBM}}(\mathscr{E}=0)$	& $E_{{\rm VBM}}(\mathscr{E}_{\rm max})$	& $E_{{\rm CBM}}(\mathscr{E}=0)$	& $E_{{\rm CBM}}(\mathscr{E}_{\rm max})$	& $\frac{\partial E_{{\rm VBM}}}{\partial \mathscr{E}}$	& $\frac{\partial E_{{\rm CBM}}}{\partial \mathscr{E}}$ \\
\vspace{-3mm}\\
\hline
Chalcogen	& $\downarrow$	& X	& X	& X	& X	& X	& $\downarrow$	& $\uparrow$/X$^{a}$\\
TM period	& X	& $\uparrow$/X$^{b}$	& $\uparrow$/X$^{c}$	& X/$\uparrow^{d}$	& $\uparrow$	& $\uparrow$	& $\uparrow$ 	& X\\
TM group	& $\downarrow$	& $\uparrow$/$\downarrow^{e}$	& $\uparrow$	& $\uparrow$	& $\uparrow$ 	& $\uparrow$	& $\downarrow$	& $\uparrow$\\
Phase	& $\uparrow$	& $\uparrow$	& $\uparrow$	& $\uparrow$	& $\uparrow$	& $\uparrow$	& $\uparrow$	& $\uparrow$\\
Layers	& $\uparrow$	& $\downarrow$	& $\uparrow$	& $\uparrow$	& $\downarrow$	& $\downarrow$& $\uparrow$	& $\downarrow$\\
\hline
\end{tabular}
\begin{tablenotes}
{\footnotesize
\item$^{a}$ 48/48\%; $^{b}$ 48/39\%; $^{c}$ 44/34\%; $^{d}$ 41/38\%; $^{e}$ 47/37\%.}
\end{tablenotes}
\caption{Multidimensional analysis of dominant behaviour(s) within the TMDC family for the calculated parameters: giant Stark coefficient $\hat{S}$, critical field for insulator-to-metal/semi-metal transition $\mathscr{E}_{F_{c}}$, VBM and CBM energies at zero and maximum (as defined in the main text) fields, and the VBM and CBM bending rates with field.  F (not exhibited), $\uparrow$, $\downarrow$, and X respectively indicate flat, monotonically increasing, monotonically decreasing, and other non-monotonic behaviour with an increase to the relevant dimension. The dimensions (with the ordering used in the analysis) are: chalcogen (period 2, 3, 4, or 5), transition-metal period (3, 4, or 5), transition-metal group (III, IV, VI, or X), material phase (H or T), and the number of atomic layers (2 or 3). Single indicators are shown when $>$50\% of the subset classes exhibit the behavior; pairs of indicators are shown when there are two behaviors that capture most subsets (their relative percentages are shown as footnotes).}
\label{tab:summary}
\end{ruledtabular}
\end{table*}

We notice an overall decrease in band gap variation under field, $\hat S$, down the chalcogen group from the oxides to tellurides with a few anomalies. This trend is also illustrated in Fig.~\ref{Fig:BandGaps} where we can see a decrease in the bar lengths from oxides (blue) to tellurides (black) for most of the transition metals, which again suggest that tellurides are less sensitive to electric fields than other dichalcogenides.

$\hat{S}$ behaves non-monotonically with transition-metal period, but decreases with transition-metal group.  The group X TMDCs exhibit the least band gap variation under field when compared as a whole to other groups. For example, the PtO$_{2}$ bilayer, the PtS$_{2}$ and the PtSe$_{2}$ bilayer and trilayer do not respond to the field within the range of 0.2 V/\r{A} as compared to the Hf- and W-based dichalcogenides (Table~\ref{Tab: Table1}).

The GSE coefficients increase from the H- to T-phase materials as shown in Fig.~\ref{Fig:BandGaps} for those materials which are stable in both H and T phases. This is mainly due to the distinct intralayer stacking of the two phases leading to a difference in the interplanar X--X dispersion interactions \cite{Zr_Field}. Thus selection of the material phase is another potential pre-fabrication lever for controlling band gaps and their variations under field.

We further reveal the critical fields ($\mathscr{E}_{F_{c}}$), where closures of band gaps occur and the bilayer or trilayer TMDCs undergo semiconductor-to-metal phase transitions. To achieve this, we increased the field strengths beyond 0.2 V/\r{A} for those materials whose band gaps had not yet closed. We report these computed critical fields in Table \ref{Tab: Table1}, and summarise their trends in Table \ref{tab:summary}. 

The critical field strengths vary non-monotonically down the chalcogen group from oxides to tellurides. Similarly, no dominant trends in the critical field strengths are predicted across the transition metal periods or down the transition metal groups for the same phase, chalcogen, and number of layers.  However, $\mathscr{E}_{F_{c}}$ does increase from H- to T-phase materials, and and is generally lower for trilayer materials than the related bilayer ones.

The converse dependencies of $\hat{S}$ and $\mathscr{E}_{F_{c}}$ on transition-metal group or layer number are understandable; since $\hat{S}$ corresponds to the slope of the band gap with field, it stands to reason that more responsive materials (trilayers) should become metallic or semi-metallic at lower fields ($\mathscr{E}_{F_{c}}$), particularly once account is taken of any discrepancy in their zero-field band gaps.  However, the behavior with phase is aligned, with both parameters increasing.  Comparison with Table \ref{tab:bulksum} shows that this is explained by an evident accompanying increase in the zero-field band gaps with phase.

We defer discussing compound trends until the rest of Table \ref{tab:summary} has been explored in Sec. \ref{sec:VCBM}.
 
Our general findings are that: trilayer band gaps respond more to field than bilayer gaps; T-phase materials respond more than H-phase ones; band gap modulations decrease from oxides to tellurides; and the response to the electric field decreases from left to right across the transition metals when compared within the same period and for the same chalcogens. These findings reveal the whole range of TMDC band gap responses under field. They enable one to select the appropriate material(s) to engineer devices in according to one's application's requirements.

\subsection{Valence and conduction band absolute positions and variations under field} 
\label{sec:VCBM}

\begin{figure*}[tb!]
\centering
\includegraphics[width=0.99\linewidth]{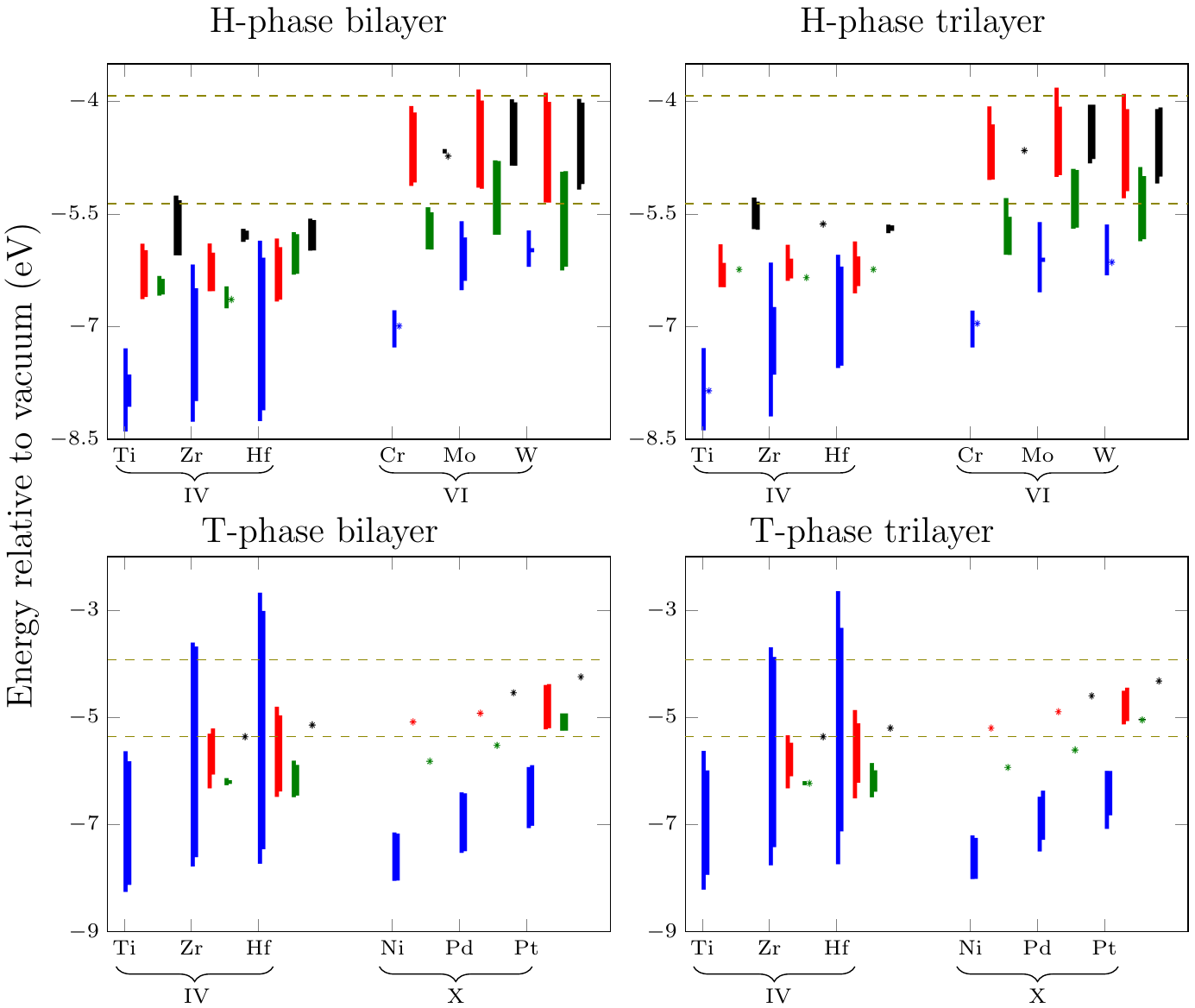}
\caption{Positions of the valence band maxima (VBMs) and conduction band minima (CBMs) with respect to the vacuum level (0 eV) in H- and T-phase, bilayer and trilayer TMDCs as labeled. The bottom of each rectangular bar shows the VBM and the top the CBM; its length shows the band gap. The color of the bars represent the chalcogens; blue for oxides, red for sulphides, green for selenides, and black for tellurides. The double bars show the band positions at zero field (left) and at the maximum field of 0.2 V/\r{A} or less (as discussed in the main text and in Table \ref{Tab: Table1}) for the materials where the band gap closes before 0.2 V/\r{A} (right). Asterisks denote the Fermi levels of zero-band gap materials at their critical fields (or at zero field for materials which are naturally metallic or semimetallic). Note: the critical fields can be found in Table \ref{Tab: Table1}. Dashed lines show the energies 0.1 eV above the hydrogen evolution potential, \textit{i.e.}, -4.03 eV and 0.1 eV below the oxygen evolution potential, \textit{i.e.}, -5.23 eV.}
\label{Fig:VBM_CBM}
\end{figure*}

For many applications, the energies of the valence and conduction bands with respect to the vacuum level are important. For example, in the construction of 2D heterostructure materials, knowledge of absolute VBM and CBM energies is required to predict behavior. For device design, such as a single-electron transistor \cite{SET_Angus, SET}, where we require control over its transport properties, study of the VBMs' and CBMs' responses to electric fields is essential.     

In Fig.~\ref{Fig:VBM_CBM}, we show the computed energies of all studied TMDCsâ VBMs, and CBMs with respect to the vacuum -- at zero field (left rectangular bar) and for the maximum field of 0.2 V/\r{A} or less for those materials whose band gaps close before 0.2 V/\r{A} (right rectangular bar). The lengths of the rectangular bars show the band gaps. The blue, red, green, and black colors show oxides, sulphides, selenides, and tellurides respectively.  

We also summarise any trends in Table \ref{tab:summary}.  Here, for the absolute energies, we have $20(\times 4)+2(\times 3)$ subset classes when varying chalcogen, $24(\times 3)+8(\times 2)$ varying transition-metal period, $42(\times 2)$ varying transition-metal group, $18(\times 2)$ varying phase, and $44(\times 2)$ varying layer number. The rates of change of the VBM and CBM have the same numbers and types of subset classes as $\hat{S}$ and $\mathscr{E}_{F_{c}}$, described in Sec. \ref{sec:bands}.

The absolute VBM and CBM energies behave non-monotonically as the chalcogen period increases.  However, they largely increase with transition-metal period and group, and with changing from H to T phase.  Whilst the zero-field extrema energies increase with the number of layers, the maximum-field extrema energies both decrease with layer number, consistent with the increased $\hat{S}$ also exhibited by trilayer materials.

When changing transition metal period, 11 of 32 subset classes show an overall increases in all four reported VBM and CBM energies (Fig.~\ref{Fig:VBM_CBM} top panels). This behavior is particularly prominent in the oxides.  28 of 42 subset classes show the same behaviour when transition-metal group is changed (\textit{i.e.}, comparing same-colored bars between left and right groups in subfigures of Fig.~\ref{Fig:VBM_CBM}).  However, while increasing the number of layers 16 of 44 subset classes show increases to the zero-field extrema with decreases to the high-field extrema.

For almost all of these materials, we find significant bending in the CBM under field (as reported in Table \ref{Tab: Table1} and shown in Fig.~\ref{Fig:VBM_CBM}) except for a few materials, \textit{e.g.}, WSe$_{2}$ bilayers and PdO$_{2}$ trilayers show anomalous behavior; their CBMs increase with the field. The VBMs for all the materials show increasing energy shifts under electric field except TiTe$_{2}$ bilayers, H-HfTe$_{2}$ bilayers, CrTe$_{2}$ bilayers, and MoS$_{2}$ bilayers, which show a negative bending in their VBMs (Table \ref{Tab: Table1}). 

Looking broadly across all parameters reported in Table \ref{tab:summary}, we identify the following large-scale compound behaviours: when varying transition-metal group, 10 of 42 subset classes show consistent behavior across all eight parameters -- increases to all absolute energies plus $\partial E_{\rm CBM}/\partial\mathscr{E}$ combined with decreases to the other three parameters; when increasing the number of layers, 19 of 44 subsets show increasing $\hat{S}$ and $\partial E_{\rm VBM}/\partial\mathscr{E}$ with decreasing $\mathscr{E}_{F_{c}}$ and $\partial E_{\rm CBM}/\partial\mathscr{E}$, 15 of 44 show increased $\hat{S}$ and VBM energies (both fields) with decreased $\mathscr{E}_{F_{c}}$, and 23 of 44 show decreased $\mathscr{E}_{F_{c}}$ with increased VBM energies (both fields).  Other compound behaviours are only exhibited by small fractions of the subset classes.

It has been reported in \cite{RT2015} that materials with CBMs above the standard hydrogen electrode (SHE), \textit{i.e.} -4.03 eV relative to vacuum at pH 7, can be used at the cathode of photocatalytic water splitting devices to evolve hydrogen. Similarly materials with VBMs below the oxygen evolution potential (1.23 eV below the SHE) can be used as photoanodes in water splitting devices. It has been suggested that the CBM/VBM should lie a few tenths of an electron volt above/below the redox potentials \cite{Trasatti1986} to account for the intrinsic energy barriers of the water splitting reactions \cite{watersplitting, Butler1978}. Any material with a CBM a few tenths of an electron volt above -4.03~eV or VBM a few tenths of an electron volt below -5.26~eV is desirable for water splitting applications. 

In Fig.~\ref{Fig:VBM_CBM}, the CBMs of Mo- and W-based, bilayer and trilayer sulphides lie $\geq$ 0.1 eV above -4.03 eV at zero and low fields. For example, MoS$_{2}$ bilayers have a useful CBM for water splitting applications in electric fields ranging from 0 to 0.14 V/\r{A}. Similarly, MoS$_{2}$ trilayers, WS$_{2}$ bilayers, and WS$_{2}$ trilayers are useful for this application in fields ranging from 0 to 0.04, 0.1, and 0.04 V/\r{A}, respectively. The T-phase, Zr and Hf, bilayer and trilayer oxides have CBMs well above the SHE potential for both zero and maximum field and could be useful for water splitting applications under any field strength from 0 to 0.2 V/\r{A}. 

Most of the materials have their VBM below the oxygen redox potential except for the bilayers and trilayers of CrS$_{2}$, CrTe$_{2}$, MoS$_{2}$, MoTe$_{2}$, WS$_{2}$, WTe$_{2}$, PtS$_{2}$, and PtSe$_{2}$, and the zero band gap materials (Fig.~\ref{Fig:VBM_CBM}). The TMDCs therefore appear to be a highly useful class of materials for water splitting.

\section{Conclusions}
We have presented a comprehensive density-functional theory study of the electronic structures of 192 configurations of 39 stable, two-dimensional, transition-metal dichalcogenides. Our calculations show the band gaps of few-layer TMDCs along with variation of the band structures by electric fields.  The data has been analysed across five dimensions: chalcogen period, transition-metal period, transition-metal group, phase, and number of layers.

Band gaps generally decrease down the chalcogen group from oxides to tellurides, increase across the transition metals from left to right, are larger for T-phase materials than corresponding H-phase ones, and decrease with more layers. The responses to the electric field decrease down the chalcogens and across transition metals in the same period, are larger for T-phase materials than H-phase ones, and increase with increasing number of layers. We generally found that the CBMs decrease with higher fields while the VBMs increase, narrowing the band gap from both sides. We also have suggested materials which could be useful for water splitting applications under zero and low fields.  

By presenting the field-modulated behavior of monolayer, bilayer, and trilayer structures of 39 different materials, this work supports the advance of 2D materials from fundamental research to real applications. In particular, it will aid band-gap and opto-electronic engineers to select the optimal TMDC for their device requirements.

 \section*{Acknowledgements}

The authors acknowledge financial support from the Australian Research Council (Project Nos. DP130104381, CE140100003, FT160100357, LE160100051, and CE170100026). This work was supported by computational resources provided by the Australian Government through the National Computational Infrastructure (NCI) under the National Computational Merit Allocation Scheme.

\bibliography{TMDCRef}

\onecolumngrid \phantom{4}
\begin{sidewaystable*}[p!]
\centering
\caption{Relaxed lattice parameters (in-plane $a$ and plane-perpendicular $c$), interlayer separation $I_{S}$, band gaps of bulk structures without electric field ($E_{g}$), a review of the published experimental and/or theoretical band gaps at zero field for various numbers of layers, our computed band gaps $E_{g}$ at zero field for various numbers of layers, a review of the published giant-Stark-effect (GSE) coefficients ($\hat S$) under electric field, our computed $\hat S$ along with the upper and lower boundaries of the 95\% confidence intervals, trilayer to bilayer $\hat S$ ratio, bending rates of the VBMs and CBMs under field, published and our computed critical fields $\mathscr{E}_{F_{c}}$ where semiconductor-to-metal/semimetal phase transitions occur, and the types of material after the phase transition (M for metal, SM for semimetal).}
\label{Tab: Table1}
\begin{tabular}{cccccccccccccccc}
\toprule[0.4mm]
\parbox[t]{1.2cm}{ Material}                   & \parbox[t]{1cm}{$a$ (\r{A})}              & \parbox[t]{1cm}{$c$ (\r{A})}             & \parbox[t]{1cm}{$I_{S}$ (\r{A})}              &  \parbox[t]{1.2cm}{Bulk $E_{g}$\\(eV)}      & \parbox[t]{1cm}{Layers} & \parbox[t]{2.8cm}{Published \\ $E_{g}$ at \\ zero field \\ (eV)}  & \parbox[t]{2cm}{$E_{g}$ at\\zero field \\(eV) }  &\parbox[t]{1.2cm}{Published\\$\hat S$ (\r{A})}  & \parbox[t]{1.2cm}{ $\hat S$\\ (\r{A})} &\parbox[t]{1cm}{ $\hat S$ ratio \\ (tri/bi)} & \parbox[t]{1.4cm}{VBM \\ bending \\(eV/V/\r{A})}&  \parbox[t]{1.4cm}{ CBM \\ bending \\(eV/V/\r{A})} &\parbox[t]{1.5cm}{Published \\ $\mathscr{E}_{F_{c}}$\\ (V/\r{A})}&\parbox[t]{1.2cm}{$\mathscr{E}_{F_{c}}$\\ (V/\r{A})}	& Type
\\
 
 \midrule[0.3mm]

\multirow{3}{*}{H-ScO$_{2}$}  & \multirow{3}{*}{3.21} & \multirow{3}{*}{8.32} & \multirow{3}{*}{4.16} & \multirow{3}{*}{0.00} &     Mono   &   \parbox[t]{2.8cm}{1.05\cite{stabilityPaper}, 1.521\cite{SOC_MLS},\\2.90\cite{ScO2}, 1.30\cite{ScO2_ML}}                                                                     & 2.59       &-        &   0.00   &  & - &-& -&-  & - \\
                &                   &                   &                   &                 &   Bi     &     0 \cite{ScO2}&   0.00            &  -&   0.00   &  - & - &- & -&- & -\\
           &                         &                   &                   &                   &      Tri  &       -&      0.00   &-    &   0.00 &  & - &- & -  &-  & - \\
\hline
\multirow{3}{*}{H-ScS$_{2}$} & \multirow{3}{*}{3.73} & \multirow{3}{*}{10.5} & \multirow{3}{*}{5.25} &\multirow{3}{*}{0.00} &     Mono   &  \parbox[t]{2.8cm}{0.44\cite{stabilityPaper}, 0.721\cite{SOC_MLS},\\0.74\cite{Scandium}}                                                                      &  1.58              &  - &  0.00 &  & - &-& -&- &-  \\
           &                          &                   &                   &                 &   Bi     &    -&   0.00            & -&   0.00 & - & - &- & -&-   &- \\
        &                            &                   &                   &                   &      Tri  &      - &       0.00      & -&    0.00 & & - &- & -& -&- \\
\hline
\multirow{3}{*}{H-ScSe$_{2}$} & \multirow{3}{*}{3.88} & \multirow{3}{*}{11.75} & \multirow{3}{*}{5.88} &\multirow{3}{*}{0.00} &     Mono   &   0.27\cite{stabilityPaper}, 0.456 \cite{SOC_MLS}                                                                               &  1.57              &  - &  0.00   &   & - &- & -&- &- \\
       &                              &                   &                   &                 &   Bi     &    -&   0.00            & -&   0.00 & - & - &- & -&-  &- \\
   &                                 &                   &                   &                   &      Tri  &      -&       0.00      & -&    0.00 & & - &- & -&-  &-\\
\hline
\multirow{3}{*}{H-ScTe$_{2}$} & \multirow{3}{*}{4.15} & \multirow{3}{*}{11.91} & \multirow{3}{*}{5.95} &\multirow{3}{*}{0.00} &     Mono   &   0.00 \cite{stabilityPaper}                                                                               &  1.43              &  - &  0.00 &  & - &-& -&-   &-  \\
   &                                 &                   &                   &                 &   Bi     &    -&   0.00            & -&   0.00  & - & -&- & -&- &-  \\
  &                                  &                   &                   &                   &      Tri  &      -&       0.00      & -&    0.00 & & - &- & -&- &-\\                  
                  
 \midrule[0.3mm]
 
\multirow{3}{*}{H-TiO$_{2}$}  & \multirow{3}{*}{2.90} & \multirow{3}{*}{13.30} & \multirow{3}{*}{6.65} &\multirow{3}{*}{1.08}  &     Mono   &   1.10 \cite{RT2015}, 2.1 \cite{H-TiO2Structure}                                                                               &  1.12              &  - &  0.00   &  & - &-&-&-  &- \\
  &                                  &                   &                   &                 &   Bi     &    -&   1.09            & -&   3.43 $\pm$ 0.02     & 2.00 & 1.69 &-1.73& -&0.32 & SM \\
 &                                   &                   &                   &                   &      Tri  &      -&       1.08      & -&    6.87 $\pm$ 0.04 & & 3.45 &-3.41 & - &0.16  & SM\\                  
 \multirow{3}{*}{T-TiO$_{2}$} & \multirow{3}{*}{2.96} & \multirow{3}{*}{4.79} &\multirow{3}{*}{4.79} & \multirow{3}{*}{2.54} &     Mono   &   2.65\cite{RT2015}                                                                               &  2.68             &  - &  0.00  &  & - &-& -&-  &-  \\
    &                                &                   &                   &                 &   Bi     &    -&   2.61            & -&   1.62 $\pm$ 0.05 & 2.08 & 0.66 &-0.96 & -&1.60 & M \\
 &                                   &                   &                   &                   &      Tri  &      -&       2.56      & -&    3.37 $\pm$ 0.04& 
& 1.51 &-1.87 & - &0.73 & M \\
\hline
\multirow{3}{*}{H-TiS$_{2}$}  & \multirow{3}{*}{3.30} & \multirow{3}{*}{12.73} & \multirow{3}{*}{6.37} &\multirow{3}{*}{0.51} &     Mono   &   0.62 \cite{RT2015}                                                                               &  1.04              &  - &  0.00     & & - &- & - &-  &- \\
   &                                 &                   &                   &                 &   Bi    &    -&   0.72            & -&   0.60 $\pm$ 0.05     & 2.15 & 0.03 &-0.57& - &1.10 & SM \\
&                                    &                   &                   &                   &      Tri  &      -&       0.56      & -&    1.29 $\pm$ 0.08& & 0.02 &-1.3 & - &0.42 & M\\
\hline
\multirow{3}{*}{H-TiSe$_{2}$}  & \multirow{3}{*}{3.80} & \multirow{3}{*}{13.40} & \multirow{3}{*}{6.70} &\multirow{3}{*}{0.00}  &     Mono   &   0.42\cite{RT2015}                                                                               &  0.42              &  - &  0.00    & & - &- & -&- &-  \\
 &                                   &                   &                   &                 &   Bi     &    -&   0.25            & -&   0.12 $\pm$ 0.01& - & 0.02 &-0.10 & -& 1.70 & SM  \\
  &                                 &                   &                   &                   &      Tri  &      -&       0.00      & -&   -& & - &- &-&- &- \\

\hline
\multirow{3}{*}{H-TiTe$_{2}$} & \multirow{3}{*}{3.72} & \multirow{3}{*}{13.84} & \multirow{3}{*}{6.92} &\multirow{3}{*}{0.26} &     Mono   &   0\cite{RT2015}, 0\cite{stabilityPaper}                                                                               &  1.11              &  - &  0.00   &  & - &- & -&-  &-  \\
  &                                  &                   &                   &                 &   Bi     &    -&   0.78            & -&   0.29 $\pm$ 0.02   & 0.83 & -0.08 &-0.38 & -&2.40 & SM\\
 &                                   &                   &                   &                   &      Tri  &      -&       0.41      & -&   0.24 $\pm$ 0.03 & & 0.00 &-0.24 & -&1.10 & SM \\  
                  
 \midrule[0.35mm]
\multirow{3}{*}{H-ZrO$_{2}$} & \multirow{3}{*}{3.12} & \multirow{3}{*}{12.24} & \multirow{3}{*}{6.12} &\multirow{3}{*}{1.54} &     Mono   &  1.59 \cite{RT2015}                                                                               &  1.62              &  - &  0.00  & & - &- &- &-  &-   \\
  &                                  &                   &                   &                 &   Bi     &    -&   2.08        & -&   3.03 $\pm$ 0.02 & 1.91 & 1.50 &-1.53 & - &0.58 & SM \\
  &                                  &                   &                   &                   &      Tri  &      -&       2.04      & -&    5.91 $\pm$ 0.55& & 2.80 &-3.10 & - &0.30 & M  \\                  
 \multirow{3}{*}{T-ZrO$_{2}$} & \multirow{3}{*}{3.26} & \multirow{3}{*}{4.81} &\multirow{3}{*}{4.81} & \multirow{3}{*}{3.96} &     Mono   &   4.37 \cite{RT2015}                                                                               &  4.41            &  - &  0.00   & & - &- & - &-  &-  \\
  &                                  &                   &                   &                 &   Bi     &   -&   4.16            & -&   1.33 $\pm$ 0.09& 2.14 & 0.84 &-0.48 & -&2.40 & SM \\
 &                                   &                   &                   &                   &      Tri &      -&       4.05     & -&    2.85 $\pm$ 0.17
&  & 1.80 &-1.05 & - &1.28  & M\\
                                     
\bottomrule[0.4mm]
\end{tabular}
\end{sidewaystable*}

\begin{sidewaystable*}[p!]
\centering
\hspace*{-0.5 in}

\begin{tabular}{cccccccccccccccc}
\toprule[0.4mm]
\parbox[t]{1.2cm}{ Material}                   & \parbox[t]{1cm}{$a$ (\r{A})}              & \parbox[t]{1cm}{$c$ (\r{A})}             & \parbox[t]{1cm}{$I_{S}$ (\r{A})}              &  \parbox[t]{1.2cm}{Bulk $E_{g}$\\(eV)}      & \parbox[t]{1cm}{Layers} & \parbox[t]{2.8cm}{Published \\ $E_{g}$ at \\ zero field \\ (eV)}  & \parbox[t]{2cm}{$E_{g}$ at\\zero field \\(eV) }  &\parbox[t]{1.2cm}{Published\\$\hat S$ (\r{A})}  & \parbox[t]{1.2cm}{ $\hat S$\\ (\r{A})} &\parbox[t]{1cm}{ $\hat S$ ratio \\ (tri/bi)} & \parbox[t]{1.4cm}{VBM \\ bending\\(eV/V/\r{A})}&  \parbox[t]{1.4cm}{ CBM\\bending\\(eV/V/\r{A})} &\parbox[t]{1.5cm}{Published \\ $\mathscr{E}_{F_{c}}$\\ (V/\r{A})}&\parbox[t]{1.2cm}{$\mathscr{E}_{F_{c}}$\\ (V/\r{A})} & Type\\
 \midrule[0.3mm]

\multirow{3}{*}{H-ZrS$_{2}$}  & \multirow{3}{*}{3.51} & \multirow{3}{*}{11.33} & \multirow{3}{*}{5.67} &\multirow{3}{*}{0.00} &     Mono   &   0.85 \cite{RT2015}                                                                               &  0.85             &  - &  0.00   &  & -&- & -&-  &-  \\
  &                                  &                   &                   &                 &   Bi     &   0.90 \cite{Zr_Field}&   0.63            & 4.61 \cite{Zr_Field}&   0.67 $\pm$ 0.04&   1.66 & 0.15 &-0.52 & 0.2$^{a}$ \cite{Zr_Field}  &1.04 & SM\\
  &                                  &                   &                   &                   &      Tri  &      -&      0.47      & -&    1.11 $\pm$ 0.06&  & 0.20 & -0.90 & - &0.34& SM \\                  
 \multirow{3}{*}{T-ZrS$_{2}$} & \multirow{3}{*}{3.62} & \multirow{3}{*}{5.85} &\multirow{3}{*}{5.85} & \multirow{3}{*}{0.74} &     Mono   &   1.03 \cite{RT2015}, 1.10 \cite{T-ZrX2_ML}                                                                               &  1.08            &  - &  0.00  &  & - &- & - &-  &-  \\
 &                                  &                   &                   &                 &   Bi     &    1.43 \cite{Zr_Field}&   1.01           & 5.53 \cite{Zr_Field}&   0.89 $\pm$ 0.07&  2.20 & 0.57 &-0.31 & 0.27$^{a}$ \cite{Zr_Field} &0.96  & SM\\
  &                                  &                   &                   &                   &      Tri  &      -&      0.96     &-&    1.96 $\pm$ 0.10
&  & 1.08 &-0.87 & - &0.46 & M \\
\hline          
  
\multirow{3}{*}{H-ZrSe$_{2}$} & \multirow{3}{*}{4.10} & \multirow{3}{*}{15.50} & \multirow{3}{*}{7.75} &\multirow{3}{*}{0.15} &     Mono   &   0.64 \cite{RT2015}                                                                               &  0.35             &  - &  0.00    &  & - &- & - &-  &-  \\
  &                                  &                   &                   &                 &   Bi     &   0.75 \cite{Zr_Field}&   0.28            & 5.2 \cite{Zr_Field}&   3.43 $\pm$ 0.06&   - & 1.36&-2.06 & 0.15$^{a}$ \cite{Zr_Field} & 0.08 & SM\\
&                                    &                   &                   &                   &      Tri &      -&      0.00      & -&    - &  & - &- & - &- &- \\                  
\multirow{3}{*}{T-ZrSe$_{2}$}  & \multirow{3}{*}{3.74} & \multirow{3}{*}{6.07} &\multirow{3}{*}{6.07} & \multirow{3}{*}{0.01} &     Mono   &   0.25 \cite{RT2015}                                                                              , 0.45 \cite{T-ZrX2_ML} &  0.24           &  - &  0.00  &  & - &-& - &- &-   \\
  &                                  &                   &                   &                 &   Bi     &  0.9 \cite{Zr_Field}&   0.12           & 5.28 \cite{Zr_Field}&   0.41 $\pm$ 0.09 & 4.63 & 0.20 &-0.20 & 0.17$^{a}$ \cite{Zr_Field} & 0.24 & SM\\
 &                                   &                   &                   &                   &      Tri  &      -&      0.06     & -&    1.90 $\pm$ 1.90 & 
& 1.14 &-0.75 & - &0.06 & M \\

 \hline

\multirow{3}{*}{H-ZrTe$_{2}$}  & \multirow{3}{*}{3.84} & \multirow{3}{*}{14.28} & \multirow{3}{*}{7.14} & \multirow{3}{*}{0.00} &     Mono   &   0.18 \cite{RT2015}                                                                               &  0.39             &  - &  0.00      &  & - &- & -& -&-  \\
&                                   &                   &                   &                 &   Bi     &   0.30 \cite{Zr_Field}&   0.20            & 4.0 \cite{Zr_Field}&   0.12 $\pm$ 0.08 &  - & 0.09 &-0.02 & 0.07$^{a}$ \cite{Zr_Field}  & 0.70 & SM \\
  &                                 &                   &                   &                   &      Tri  &      -&      0.00      & -&    - &   & - &- & - & - &- \\                  
\multirow{3}{*}{T-ZrTe$_{2}$}  & \multirow{3}{*}{3.87} & \multirow{3}{*}{6.52} & \multirow{3}{*}{6.52}& \multirow{3}{*}{0.00} &     Mono   &   0\cite{RT2015}, 0 \cite{T-ZrX2_ML}                                                                               &  0.00           &  - &  -  &   & - &- & - &-  &-   \\
   &                                  &                   &                   &                 &   Bi     &    0.03 \cite{Zr_Field}&   0.00           & 0.6 \cite{Zr_Field}&   -  & - & - &- & 0.05$^{a}$ \cite{Zr_Field} & - &-\\
 &                                   &                   &                   &                   &      Tri &      -&      0.00     & -&  -&   & - &- & - &-&-
\\     
 \hline
 
\multirow{3}{*}{H-HfO$_{2}$}  & \multirow{3}{*}{3.10} & \multirow{3}{*}{12.00} & \multirow{3}{*}{6.00} & \multirow{3}{*}{2.00} &     Mono   &   1.80 \cite{RT2015}                                                                               &  2.11              &  - &  0.00   &   & - &- & - & - &-   \\
&                                    &                   &                   &                 &   Bi     &    -&   2.39            & -&   1.91 $\pm$ 0.05&  0.50& 0.77 &-1.10 & - &1.60  & SM\\
&                                    &                   &                   &                   &      Tri  &      -&       1.50      & -&    0.96 $\pm$ 0.02&   & 0.12 &-0.83 & - &1.32 & SM  \\                  
\multirow{3}{*}{T-HfO$_{2}$}  & \multirow{3}{*}{3.25} & \multirow{3}{*}{5.90} & \multirow{3}{*}{5.90}& \multirow{3}{*}{4.67} &     Mono   &   4.63\cite{RT2015}                                                                               &  5.00            &  - &  0.00   &   & - &- & - &-   &-  \\
&                                     &                   &                   &                 &   Bi     &    -&  5.02            & -&   3.14 $\pm$ 0.03& 2.03 & 1.50 &-1.65 & - &1.58 & M \\
&                                    &                   &                   &                   &      Tri  &      -&       5.05     & -&    6.39 $\pm$ 0.03
&  & 3.14 & -3.25 & -  &0.80 & M\\
\hline                  
  
\multirow{3}{*}{H-HfS$_{2}$}  & \multirow{3}{*}{3.48} & \multirow{3}{*}{12.83} & \multirow{3}{*}{6.42}& \multirow{3}{*}{0.43} &     Mono   &   0.93 \cite{RT2015}                                                                               &  1.07            &  - &  0.00   &  & - &- & - &-   &- \\
 &                                    &                   &                   &                 &   Bi     &    -&   0.84            & -&   0.76 $\pm$ 0.06&  2.10 & 0.19 &-0.57& - & 1.05 & SM\\
&                                     &                   &                   &                   &      Tri  &      -&      0.69      & -&    1.60 $\pm$ 0.11&   & 0.44 &-1.17 & - & 0.30& M \\                  
\multirow{3}{*}{T-HfS$_{2}$} & \multirow{3}{*}{3.63} & \multirow{3}{*}{5.85} & \multirow{3}{*}{5.85}& \multirow{3}{*}{1.30} &     Mono   &   1.06\cite{RT2015}                                                                              , 1.27 \cite{T-ZrX2_ML} &  1.71           &  - &  0.00  &   & - &- & - & -  &-    \\
&                                     &                   &                   &                 &   Bi     &    -&   1.65           & -&   1.38 $\pm$ 0.06 & 2.10 & 0.72 &-0.65 & - &1.15 & SM\\
&                                     &                   &                   &                   &      Tri  &      -&      1.62     & -&    2.89 $\pm$ 0.38
&   & 1.55 &-1.34 & - &0.55 & SM\\
\hline          
  
\multirow{3}{*}{H-HfSe$_{2}$}  & \multirow{3}{*}{4.00} & \multirow{3}{*}{14.90} & \multirow{3}{*}{7.45}& \multirow{3}{*}{0.57} &     Mono   &   0.70\cite{RT2015}                                                                               &  0.85            & - &  0.00   &   & - &-& -  & -  &-  \\
                 &                   &                   &                   &                 &   Bi     &    -&   0.55            & -&   0.20 $\pm$ 0.03& - & 0.05 &-0.15 & - &1.80 & SM \\
&                                     &                   &                   &                   &      Tri  &      -&      0.00      & -&    - &  & - &- & -&-&- \\

 \multirow{3}{*}{T-HfSe$_{2}$} & \multirow{3}{*}{3.74} & \multirow{3}{*}{6.15} & \multirow{3}{*}{6.15}& \multirow{3}{*}{0.51} &     Mono   &  0.30 \cite{RT2015}                                                                              , 0.61 \cite{T-ZrX2_ML} &  0.74           &  - &  0.00       &   & - &- &- &- &-\\
 &                                   &                   &                   &                 &   Bi     &    -&   0.65           & -&   0.64 $\pm$ 0.33 & 1.67  & 0.37 &-0.26 &- &0.88 & SM \\
&                                    &                   &                   &                   &      Tri  &      -&      0.61     & -&    1.06 $\pm$ 0.61
&   & 0.51 &-0.54 & - &0.40 & SM\\

\bottomrule[0.4mm]
\end{tabular}
\footnotetext[1]{Values estimated from the figures.}
\end{sidewaystable*}

\begin{sidewaystable*}[p!]
\centering
\hspace*{-0.5 in}

\begin{tabular}{cccccccccccccccc}
\toprule[0.4mm]
\parbox[t]{1.2cm}{ Material}                   & \parbox[t]{0.9cm}{$a$ (\r{A})}              & \parbox[t]{0.9cm}{$c$ (\r{A})}             & \parbox[t]{0.9cm}{$I_{S}$ (\r{A})}              &  \parbox[t]{1.2cm}{Bulk $E_{g}$\\(eV)}      & \parbox[t]{1cm}{Layers} & \parbox[t]{2.8cm}{Published \\ $E_{g}$ at \\ zero field \\ (eV)}  & \parbox[t]{1.7cm}{$E_{g}$ at\\zero field \\(eV) }  &\parbox[t]{1.2cm}{Published\\ $\hat S$ (\r{A})}  & \parbox[t]{1.2cm}{ $\hat S$\\ (\r{A})} &\parbox[t]{0.9cm}{ $\hat S$ ratio \\ (tri/bi)} & \parbox[t]{1.4cm}{VBM \\ bending\\(eV/V/\r{A})}&  \parbox[t]{1.4cm}{ CBM\\bending\\(eV/V/\r{A})} &\parbox[t]{1.2cm}{Published\\$\mathscr{E}_{F_{c}}$\\ (V/\r{A})}&\parbox[t]{1cm}{$\mathscr{E}_{F_{c}}$\\ (V/\r{A})} & Type\\
 \midrule[0.3mm]

\multirow{3}{*}{H-HfTe$_{2}$}  & \multirow{3}{*}{3.84} & \multirow{3}{*}{13.89} & \multirow{3}{*}{6.95}& \multirow{3}{*}{0.00} &     Mono   &   0.06 \cite{RT2015}                                                                               &  0.67             &  - &  0.00    &  & -&- & - &-  &-  \\
 &                                   &                   &                   &                 &   Bi     &    -&   0.45            & -&   0.18 $\pm$ 0.03&  1.11 & -0.37 &-0.55 & - &1.19 & M \\
&                                    &                   &                   &                   &      Tri  &      -&      0.12      & -&    0.20 $\pm$ 0.03 &   & 0.05 & -0.15 & - & 0.38 & SM\\                  
 \multirow{3}{*}{T-HfTe$_{2}$} & \multirow{3}{*}{3.95} & \multirow{3}{*}{6.67} & \multirow{3}{*}{6.67}& \multirow{3}{*}{0.00} &     Mono   &   0 \cite{RT2015}, 0 \cite{T-ZrX2_ML}                                                                               &  0.00           &  - &  -  &  & - &- & - &-  &-   \\
 &                                    &                   &                   &                 &   Bi     &    -&   0.00           & -&   - & - & - &- & -&- &-\\
&                                     &                   &                   &                   &      Tri  &      -&      0.00     & -&  -
&   & - &- & - &- &-\\                   
 
\hline

\multirow{3}{*}{H-CrO$_{2}$}  & \multirow{3}{*}{2.58} & \multirow{3}{*}{14.37} & \multirow{3}{*}{7.19}& \multirow{3}{*}{0.45} &     Mono   &  \parbox[t]{2.8cm}{ 0.50 \cite{stabilityPaper}, 0.43 \cite{RT2015},\\ 0.381 \cite{SOC_MLS}                                                                              } &  0.48              &  - &  0.00    &  & - &- & - &-   &- \\
&                                    &                   &                   &                 &   Bi     &    -&   0.47            & -&   3.91 $\pm$ 0.01&  2.00 & 2.03&-1.88 & - & 0.12  & SM\\
 &                                    &                   &                   &                   &      Tri  &      -&     0.47      & -&    7.82 $\pm$ 0.07&  & 3.83 &-4.00 & - &0.06 & SM \\                  
 
\hline
\multirow{3}{*}{H-CrS$_{2}$}  & \multirow{3}{*}{2.90} & \multirow{3}{*}{12.31} & \multirow{3}{*}{6.15}& \multirow{3}{*}{1.10} &     Mono   & \parbox[t]{2.8cm}{1.07 \cite{stabilityPaper}, 0.90 \cite{RT2015},\\ 0.929 \cite{SOC_MLS}, 0.92 \cite{T-ZrX2_ML}}                                                                               &  1.34              &  - &  0.00     &  & - &- &- &- &-  \\
&                                     &                   &                   &                 &   Bi     &    -&   1.04            & -& 0.66 $\pm$ 0.01     & 2.00  & 0.09 &-0.57 & - &1.28 & SM \\
 &                                    &                   &                   &                   &      Tri  &      -&       0.96      & -& 1.33 $\pm$ 0.05 &   & 0.08 &-1.24  & - & 0.62 & M \\
\hline
\multirow{3}{*}{H-CrSe$_{2}$}  & \multirow{3}{*}{3.08} & \multirow{3}{*}{12.77} & \multirow{3}{*}{6.39}& \multirow{3}{*}{0.67} &     Mono   & \parbox[t]{2.8cm}{0.86 \cite{stabilityPaper}, 0.70 \cite{RT2015},\\ 0.756\cite{SOC_MLS}, 0.74 \cite{T-ZrX2_ML}}                                                                               &  0.83              &  - &  0.00    &   & - &- & - &-  &- \\
                  &                   &                   &                   &                 &   Bi     &    -&   0.54           & -&   0.36 $\pm$ 0.00 &   3.57   & 0.00 &-0.36  & - &1.28 & M \\
&                                     &                   &                   &                   &      Tri  &      -&       0.74      & -&   1.27 $\pm$ 0.06 &   & 0.36 &-0.90 & - &0.54 & M \\

\hline
\multirow{3}{*}{H-CrTe$_{2}$}  & \multirow{3}{*}{3.46} & \multirow{3}{*}{12.96} & \multirow{3}{*}{6.48}& \multirow{3}{*}{0.05} &     Mono   & \parbox[t]{2.8cm}{0.60 \cite{stabilityPaper}, 0.45 \cite{RT2015},\\ 0.534 \cite{SOC_MLS}, 0.60 \cite{T-ZrX2_ML}                                                                              } &  0.34              &  - &  0.00   &   & - &-& - &-  &-  \\
&                                     &                   &                   &                 &   Bi     &    -&   0.04            & -& 0.28 $\pm$ 0.02&  -  & -0.24 &-0.52 & - &0.14$  $ & SM\\
&                                     &                   &                   &                   &      Tri  &      -&       0.00      & -&   - &   & -&- & - &- &-\\  
                  
\midrule[0.3mm]
 
\multirow{3}{*}{H-MoO$_{2}$}  & \multirow{3}{*}{2.80} & \multirow{3}{*}{12.13} & \multirow{3}{*}{6.07}& \multirow{3}{*}{0.91} &     Mono   &  \parbox[t]{2.8cm}{0.97\cite{stabilityPaper}, 0.91 \cite{RT2015}, \\0.898 \cite{SOC_MLS}                                                                              } &  1.02             &  - &  0.00     &  & - &-& - &- &-  \\
&                                    &                   &                   &                 &   Bi     &    -&   0.90            & -& 1.75 $\pm$ 0.05& 2.58 & 0.72 &-1.03 & - &0.50 & SM  \\
&                                     &                   &                   &                   &      Tri  &      -&     0.92      & -&4.51 $\pm$ 0.13&   & 2.05 &-2.27 & - &0.24 & SM\\                  
 
\hline
\multirow{3}{*}{H-MoS$_{2}$}  & \multirow{3}{*}{3.15} & \multirow{3}{*}{12.32} & \multirow{3}{*}{6.16}& \multirow{3}{*}{1.16} &     Mono   & \parbox[t]{2.8cm}{1.85\cite{EmergingPL}, 1.88 \cite{PRL105},\\1.87 \cite{stabilityPaper}, 1.58 \cite{RT2015},\\1.706\cite{SOC_MLS}, 1.9 \cite{MoWS2}                                                                              } &  1.80              &  0 \cite{RNT2011} &  0.00   &  & - &- & - &-  &-  \\
&                                    &                   &                   &                 &   Bi     & \parbox[t]{2.8cm}{1.58 \cite{PRL105}, 1.25$^{a}$ \cite{LayeredTMDC},\\1.25 \cite{RNT2011}, 1.5 \cite{MoWS2}} &   1.30            & \parbox[t]{2cm}{5.50 \cite{RNT2011}, 0.66$^{a}$\cite{LLGC2012},\\ 2.25$^{a}$ \cite{Nguyen2016}, 0.83$^{a}$ \cite{LayeredTMDC}}& 0.80 $\pm$ 0.52&  1.90 & -0.08 &-0.88 & \parbox[t]{2.2cm}{1.5$^{a}$ \cite{LayeredTMDC}, \\ 0.3 \cite{RNT2011}\\ 1.5$^{a}$ \cite{LLGC2012}, \\ 0.55 \cite{Nguyen2016}} &1.45 & SM\\
 &                                    &                   &                   &                   &      Tri  &   1.45 \cite{PRL105}, 0.7$^{a}$ \cite{MoS2_ML_EF}   &       1.18      & -6.2 \cite{MoS2_ML_EF} & 1.52 $\pm$ 0.21 &   & 0.23 &-1.29 & 0.15 \cite{MoS2_ML_EF} &0.67 & SM\\

\hline

\multirow{3}{*}{H-MoSe$_{2}$}  & \multirow{3}{*}{3.28} & \multirow{3}{*}{12.90} & \multirow{3}{*}{6.45}& \multirow{3}{*}{0.91} &     Mono   & \parbox[t]{2.8cm}{1.32 \cite{RT2015}, 1.62\cite{stabilityPaper},\\ 1.438 \cite{SOC_MLS}                                                                              } &  1.43              &  0 \cite{RNT2011} &  0.00    &   & - &-& - &-   &- \\
&                                    &                   &                   &                 &   Bi     &   1.14 \cite{RNT2011} &   0.98           & 6.25 \cite{RNT2011}, \cite{LayeredTMDC} &   0.06 $\pm$ 0.01 & 3.67  & 0.00 &-0.06 & \parbox[t]{2.2cm}{1.5$^{a}$ \cite{LayeredTMDC},\\ 0.25 \cite{RNT2011}} & 2.50 & M \\
 &                                    &                   &                   &                   &      Tri  &      -&       0.78      &-& 0.23 $\pm$  0.04&  & 0.17 &-0.06 & - & 1.50 & M  \\
\bottomrule[0.4mm]
\end{tabular}

\footnotetext[1]{Values estimated from the figures.}
\end{sidewaystable*}

\begin{sidewaystable*}[p!]
\centering
\begin{tabular}{cccccccccccccccc}
\toprule[0.4mm]
\parbox[t]{1.2cm}{ Material}                   & \parbox[t]{0.9cm}{$a$ (\r{A})}              & \parbox[t]{0.9cm}{$c$ (\r{A})}             & \parbox[t]{0.9cm}{$I_{S}$ (\r{A})}              &  \parbox[t]{1.2cm}{Bulk $E_{g}$\\(eV)}      & \parbox[t]{1cm}{Layers} & \parbox[t]{2.8cm}{Published \\ $E_{g}$ at \\ zero field \\ (eV)}  & \parbox[t]{2cm}{$E_{g}$ at\\zero field \\(eV) }  &\parbox[t]{1.0cm}{Published\\$\hat S$ (\r{A})}  & \parbox[t]{1cm}{ $\hat S$\\ (\r{A})} &\parbox[t]{1cm}{ $\hat S$ ratio \\ (tri/bi)} & \parbox[t]{1.4cm}{VBM \\ bending\\(eV/V/\r{A})}&  \parbox[t]{1.4cm}{ CBM \\ bending\\(eV/V/\r{A})} &\parbox[t]{1.5cm}{Published \\ $\mathscr{E}_{F_{c}}$\\ (V/\r{A})}&\parbox[t]{1.2cm}{$\mathscr{E}_{F_{c}}$\\ (V/\r{A})}& Type \\
 \midrule[0.3mm] 

\multirow{3}{*}{H-MoTe$_{2}$} &  \multirow{3}{*}{3.51} & \multirow{3}{*}{13.97} &\multirow{3}{*}{6.99} & \multirow{3}{*}{0.80} &     Mono   &\parbox[t]{2.8cm}{1.25 \cite{stabilityPaper}, 0.93 \cite{RT2015},\\ 1.116\cite{SOC_MLS}}                                                                               &  1.09             &  - &  0.00    &  & - &- & - &  - &-  \\
 &                                   &                   &                   &                 & Bi  &  0.90 \cite{RNT2011}&   0.90            & 6.62 \cite{RNT2011}& 0.09 $\pm$ 0.09& 3.31 & 0.04 &-0.05 & 0.2 \cite{RNT2011} &1.79 & M \\
&                                    &                   &                   &                   &      Tri  &     -&       0.76      & -&0.31 $\pm$ 0.03 &   & 0.23 &-0.07 & - &0.85 & M \\

 \midrule[0.3mm]
\multirow{3}{*}{H-WO$_{2}$} & \multirow{3}{*}{2.80} & \multirow{3}{*}{12.15} & \multirow{3}{*}{6.08} &\multirow{3}{*}{0.70} &     Mono   &  \parbox[t]{2.8cm}{1.37\cite{stabilityPaper}, 1.32 \cite{RT2015},\\ 1.349\cite{SOC_MLS}                                                                              } &  0.79              & - &  0.00  &  & - &- & - &-   &-  \\
 &                                    &                   &                   &                 &   Bi     &    -&   0.47            & -& 2.24 $\pm$ 0.06&  1.68 & 1.04 &-1.20 & - &0.21 & SM \\
 &       &                   &                   &                   &      Tri  &  - &  0.67  & -&3.77 $\pm$ 0.20&  & 1.55 &-2.22 & - &0.18 & SM\\                  
 
\hline                  
  
\multirow{3}{*}{H-WS$_{2}$}  & \multirow{3}{*}{3.15} & \multirow{3}{*}{12.32} &\multirow{3}{*}{6.16} & \multirow{3}{*}{1.30} &     Mono   &  \parbox[t]{2.8cm}{1.51 \cite{RT2015}, 1.98\cite{stabilityPaper},\\ 2.1\cite{MoWS2}, 1.771 \cite{SOC_MLS}                                                                              } &  1.90           &  - &  0.00     &  & - &- & - &- &-  \\
&                                     &                   &                   &                 &   Bi     &1.36\cite{RNT2011}, 1.5 \cite{MoWS2}&   1.44            & \parbox[t]{2cm}{5.91 \cite{RNT2011}, 0.90$^{a}$ \cite{LayeredTMDC},\\ 5.83$^{a}$\cite{WS2BL}}& 0.62 $\pm$ 0.07 & 2.50 & 0.01 &-0.60& \parbox[t]{2cm}{0.27 \cite{RNT2011}, 1.2$^{a}$ \cite{LayeredTMDC},\\ 0.233 \cite{WS2BL}} &1.90 & M \\
 &                                    &                   &                   &                   &      Tri  &      -&      1.38      & -& 1.55 $\pm$ 0.65 &   & 0.27 &-1.28  & - &0.82 & SM \\                  
 
\hline          
  
\multirow{3}{*}{H-WSe$_{2}$}  & \multirow{3}{*}{3.28} & \multirow{3}{*}{12.96} &\multirow{3}{*}{6.48} & \multirow{3}{*}{0.92} &     Mono   &   \parbox[t]{2.8cm}{1.22\cite{RT2015}, 1.68\cite{stabilityPaper},\\1.535 \cite{SOC_MLS}                                                                             }  &  1.50            &  - &  0.00    &  & - &- & - &-  &-   \\
&                                     &                   &                   &                 &   Bi     &    1.2$^{a}$ \cite{WSe2MultiL}&   1.40            & \parbox[t]{2cm}{ 0.84$^{a}$\cite{LayeredTMDC},\\6$^{a}$ \cite{WSe2MultiL}} & 0.23 $\pm$ 0.07 &   3.30 & 0.24 &0.00 & 1.24\cite{LayeredTMDC}, 0.2\cite{WSe2MultiL} &2.70 & SM \\
 &                                    &                   &                   &                   &      Tri  &    1.1$^{a}$ \cite{WSe2MultiL} &      0.97      & 11.34$^{a}$ \cite{WSe2MultiL}& 0.77 $\pm$ 0.07 &    & 0.19 &-0.58 & 0.097\cite{WSe2MultiL} &1.04 & SM \\                  
 
\hline          
  
\multirow{3}{*}{H-WTe$_{2}$}  & \multirow{3}{*}{3.46} & \multirow{3}{*}{14.28}& \multirow{3}{*}{7.14} & \multirow{3}{*}{1.10} &     Mono   &   1.24\cite{stabilityPaper} &  1.28             &  - &  0     .00 &   &- &-& - &-&-  \\
&                                     &                   &                   &                 &   Bi     &   - &   1.20            & -& 0.58 $\pm$ 0.06  &  0.76& 0.35 &-0.22 & - &1.80 & M \\
 &                                    &                   &                   &                   &      Tri  &      -&      1.00      & -&  0.44 $\pm$ 0.11 &   & 0.39 &-0.05 & - &0.90 & M \\                        

\midrule[0.3mm]

\multirow{3}{*}{T-NiO$_{2}$}  & \multirow{3}{*}{2.80} & \multirow{3}{*}{4.42} & \multirow{3}{*}{4.42}& \multirow{3}{*}{0.89} &     Mono   &  \parbox[t]{2.8cm}{1.17 \cite{RT2015}, 1.38\cite{stabilityPaper},\\ 1.265\cite{SOC_MLS}                                                                              } &  1.18              &  - &  0.00    &  &- &- & - &-  &- \\
&                                     &                   &                   &                 &   Bi     &    -&   0.88            & -&   0.15 $\pm$ 0.02& 2.88 &0.01 &-0.13 & - &4.0 & SM \\
&                                     &                   &                   &                   &      Tri  &      -&     0.80      & -& 0.42 $\pm$ 0.04 &   &0.12 &-0.28 & - & $> 0.20^{b}$ & M \\                  
  \hline
  
  \multirow{3}{*}{T-NiS$_{2}$}  & \multirow{3}{*}{3.28} & \multirow{3}{*}{4.46} & \multirow{3}{*}{4.46}& \multirow{3}{*}{0.00} &     Mono   &   \parbox[t]{2.8cm}{0.51 \cite{RT2015}, 0.51\cite{stabilityPaper},\\ 0.561\cite{SOC_MLS}, 0.51 \cite{NiPdPtX2BL}}                                                                               & 0.54             &  - &  0.00       &    & - &- & - &- &-\\
&                                     &                   &                   &                 &   Bi     &    0 \cite{NiPdPtX2BL} &   0.00            & -&   -  &-  & - &- & - &-  &- \\
&                                     &                   &                   &                   &      Tri  &      -&      0.00      & -&    - &   & - &- & - &- &-\\
\hline
\multirow{3}{*}{T-NiSe$_{2}$}  & \multirow{3}{*}{3.55} & \multirow{3}{*}{5.16} & \multirow{3}{*}{5.16}& \multirow{3}{*}{0.00} &     Mono   &  \parbox[t]{2.8cm}{0 \cite{RT2015},0.10 \cite{stabilityPaper},\\0.094\cite{SOC_MLS}}                                                                               &  0.25            &  - &  0.00    & & - &- & - &-   &- \\
 &                                    &                   &                   &                 &   Bi     &    -&   0.00           & -&   - & - & - &- & - &-&-\\
&                                     &                   &                   &                   &      Tri  &      -&       0.00      & -&   - &  & -&- & - &- &-\\

\bottomrule[0.4mm]
\end{tabular}
    \footnotetext[1]{ Values estimated from the figures.}
        \footnotetext[2]{ Structure does not converge at high fields.}

\end{sidewaystable*}

\begin{sidewaystable*}[p!]
\centering

\begin{tabular}{cccccccccccccccc}
\toprule[0.4mm]
\parbox[t]{1.2cm}{ Material}                   & \parbox[t]{1cm}{$a$ (\r{A})}              & \parbox[t]{1cm}{$c$ (\r{A})}             & \parbox[t]{1cm}{$I_{S}$ (\r{A})}              &  \parbox[t]{1.2cm}{Bulk $E_{g}$\\(eV)}      & \parbox[t]{1cm}{Layers} & \parbox[t]{2.8cm}{Published \\ $E_{g}$ at \\ zero field \\ (eV)}  & \parbox[t]{2cm}{$E_{g}$ at\\zero field \\(eV) }  &\parbox[t]{1.2cm}{Published\\$\hat S$ (\r{A})}  & \parbox[t]{1.2cm}{ $\hat S$\\ (\r{A})} &\parbox[t]{1.2cm}{ $\hat S$ ratio \\ (tri/bi)} & \parbox[t]{1.4cm}{VBM \\ bending\\(eV/V/\r{A})}&  \parbox[t]{1.4cm}{ CBM \\ bending\\(eV/V/\r{A})} &\parbox[t]{1.5cm}{Published \\ $\mathscr{E}_{F_{c}}$\\ (V/\r{A})}&\parbox[t]{1.2cm}{$\mathscr{E}_{F_{c}}$\\ (V/\r{A})} & Type \\
 \midrule[0.3mm]

\multirow{3}{*}{T-PdO$_{2}$}  & \multirow{3}{*}{3.07} & \multirow{3}{*}{4.26} & \multirow{3}{*}{4.26}& \multirow{3}{*}{1.12} &     Mono   &   1.30 \cite{RT2015}                                                                               &  1.39            &  - &  0.00     &  & - &- & - &- &- \\
&                                    &                   &                   &                 &   Bi     &    -&  1.10           & -& 0.25 $\pm$ 0.03 &  2.52 & 0.20 &-0.05 & - &2.50 & SM \\
&                                     &                   &                   &                   &      Tri  &      -&     1.00      & -& 0.63 $\pm$ 0.06&   & 0.92 &0.28 & - &0.98 & M \\                  
 
\hline
\multirow{3}{*}{T-PdS$_{2}$}  & \multirow{3}{*}{3.52} & \multirow{3}{*}{4.52} & \multirow{3}{*}{4.52}& \multirow{3}{*}{0.00} &     Mono   &   1.11 \cite{RT2015}, 1.11\cite{NiPdPtX2BL}                                                                               &  1.36              &  - &  0.00    &   & - &-& - &-  &- \\
&                                     &                   &                   &                 &   Bi     &    0.00 \cite{NiPdPtX2BL} &   0.01            & -&   0.00 &   - & - &- & - &- &-\\
&                                     &                   &                   &                   &      Tri  &      -&       0.00      & -&    -  &  & - &- & - &- &-\\
\hline
\multirow{3}{*}{T-PdSe$_{2}$}  & \multirow{3}{*}{3.75} & \multirow{3}{*}{4.57} & \multirow{3}{*}{4.57}& \multirow{3}{*}{0.00} &     Mono   &   0.48 \cite{RT2015}                                                                               &  0.84              &  - &  0.00  &  & - &- & -&-  &-   \\
&                                     &                   &                   &                 &   Bi     &    -&  0.00           & -&   - & - & - &- & - &-&- \\
 &                                    &                   &                   &                   &      Tri  &      -&      0.00      & -&   - &   & - &- & - &- &-\\

\hline
\multirow{3}{*}{T-PdTe$_{2}$}  & \multirow{3}{*}{4.02} & \multirow{3}{*}{5.19} & \multirow{3}{*}{5.19}& \multirow{3}{*}{0.00} &     Mono   &  0 \cite{RT2015}                                                                               &  0.30            &  - &  0.00   &   & - &- & - &-   &- \\
 &                                    &                   &                   &                 &   Bi     &    -&   0.00            & -&  - & - & - &- & - &- &-\\
&                                     &                   &                   &                   &      Tri  &      -&       0.00      & -&   - &  & - &- & - &-&- \\ 
 
  \midrule[0.3mm]
 
\multirow{3}{*}{T-PtO$_{2}$}  & \multirow{3}{*}{3.25} & \multirow{3}{*}{4.21} & \multirow{3}{*}{4.21}& \multirow{3}{*}{1.17} &     Mono   &   1.60 \cite{RT2015}                                                                               &  1.29             &  - &  0.00   &  & - &-& - &-   &- \\
&                                     &                   &                   &                 &   Bi     &    -& 1.12           & -&   0.00 &  -& - &- & -&-  &-\\
 &                                    &                   &                   &                   &      Tri  &      -&     1.06      & -& 1.30 $\pm$ 0.07&  & 1.00 &-0.30 & - &0.80 & M \\                  
 
\hline
\multirow{3}{*}{T-PtS$_{2}$} & \multirow{3}{*}{3.54} & \multirow{3}{*}{5.04} & \multirow{3}{*}{5.04}& \multirow{3}{*}{0.16} &     Mono  &\parbox[t]{2.8cm}{  1.61 \cite{RT2015}, 1.75\cite{NiPdPtX2BL}                                                                              } &  1.77              &  - &  0.00   &  & - &- & - &-&-    \\
 &                                    &                   &                   &                 &   Bi     &    0.64 \cite{NiPdPtX2BL} &  0.81           & -&   0.00 &   - & - &- & - &- &-\\
&                                     &                   &                   &                   &      Tri  &      -&       0.61      & -&    0.00 &   &- &- & - &-&- \\
\hline
\multirow{3}{*}{T-PtSe$_{2}$}  & \multirow{3}{*}{3.72} & \multirow{3}{*}{5.08} & \multirow{3}{*}{5.08}& \multirow{3}{*}{0.00} &     Mono   &   1.07 \cite{RT2015}                                                                               &  1.43              &  - &  0.00   &   & - &- & - &-   &- \\
 &                                    &                   &                   &                 &   Bi     &    -&   0.30           & -&   0.00  & - & - &- & - &-&- \\
&                                    &                   &                   &                   &      Tri  &      -&       0.01      & -&   0.00 &  & - &- & - & - &-\\

\hline
\multirow{3}{*}{T-PtTe$_{2}$}  & \multirow{3}{*}{4.02} & \multirow{3}{*}{5.22} & \multirow{3}{*}{5.22}& \multirow{3}{*}{0.00} &     Mono  &   0.23 \cite{RT2015}                                                                               &  0.83             &  - &  0.00   &  & - &- & - &-  &-  \\
&                                     &                   &                   &                 &   Bi     &    -&   0.00            & -&   - &  -& - &- &-&- &-\\
&                                     &                   &                   &                   &      Tri  &      -&       0.00      & -&   - &  & - &- & -&- &-\\ 
                                      
\bottomrule[0.4mm]
\end{tabular}
\end{sidewaystable*}
\twocolumngrid

\end{document}